\documentstyle[twoside,epsf]{article}

\catcode`\@=11
\long\def\@makefntext#1{
\protect\noindent \hbox to 3.2pt {\hskip-.9pt  
$^{{\eightrm\@thefnmark}}$\hfil}#1\hfill}		

\def\thefootnote{\fnsymbol{footnote}}
\def\@makefnmark{\hbox to 0pt{$^{\@thefnmark}$\hss}}	
	
\def\ps@myheadings{\let\@mkboth\@gobbletwo
\def\@oddhead{\hbox{}
\rightmark\hfil\eightrm\thepage}   
\def\@oddfoot{}\def\@evenhead{\eightrm\thepage\hfil
\leftmark\hbox{}}\def\@evenfoot{}
\def\sectionmark##1{}\def\subsectionmark##1{}}

\oddsidemargin=\evensidemargin
\renewcommand{\thefootnote}{\fnsymbol{footnote}}

\newcounter{sectionc}\newcounter{subsectionc}\newcounter{subsubsectionc}

\def\thesectionc       {\arabic{sectionc}}
\def\thesubsectionc    {\thesectionc.\arabic{subsectionc}}
\def\thesubsubsectionc 
                 {\thesectionc.\arabic{subsectionc}.\arabic{subsubsectionc}}

\renewcommand{\section}[1] {
        \vspace{12pt}\refstepcounter{sectionc}
	\setcounter{subsectionc}{0}\setcounter{subsubsectionc}{0}
	\noindent{\tenbf\thesectionc. #1}\par\vspace{5pt}}

\renewcommand{\subsection}[1] {
	\vspace{12pt}\refstepcounter{subsectionc}
	\setcounter{subsubsectionc}{0}
	\noindent{\bf\thesubsectionc. {\kern1pt \bfit #1}}\par\vspace{5pt}}

\renewcommand{\subsubsection}[1] {
	\vspace{12pt}\refstepcounter{subsubsectionc}
	\noindent{\tenrm\thesubsubsectionc. 
        {\kern1pt \tenit #1}}\par\vspace{5pt}}

\newcommand{\nonumsection}[1] {\vspace{12pt}\noindent{\tenbf #1}
	\par\vspace{5pt}}

\newcounter{appendixc}
\newcounter{subappendixc}[appendixc]
\newcounter{subsubappendixc}[subappendixc]
\renewcommand{\thesubappendixc}{\Alph{appendixc}.\arabic{subappendixc}}
\renewcommand{\thesubsubappendixc}
	{\Alph{appendixc}.\arabic{subappendixc}.\arabic{subsubappendixc}}

\renewcommand{\appendix}[1] {\vspace{12pt}
        \refstepcounter{appendixc}
        \setcounter{figure}{0}
        \setcounter{table}{0}
        \setcounter{lemma}{0}
        \setcounter{theorem}{0}
        \setcounter{corollary}{0}
        \setcounter{definition}{0}
        \setcounter{equation}{0}
        \renewcommand{\thefigure}{\Alph{appendixc}.\arabic{figure}}
        \renewcommand{\thetable}{\Alph{appendixc}.\arabic{table}}
        \renewcommand{\theappendixc}{\Alph{appendixc}}
        \renewcommand{\thelemma}{\Alph{appendixc}.\arabic{lemma}}
        \renewcommand{\thetheorem}{\Alph{appendixc}.\arabic{theorem}}
        \renewcommand{\thedefinition}{\Alph{appendixc}.\arabic{definition}}
        \renewcommand{\thecorollary}{\Alph{appendixc}.\arabic{corollary}}
        \renewcommand{\theequation}{\Alph{appendixc}.\arabic{equation}}
        \noindent{\tenbf Appendix \theappendixc #1}\par\vspace{5pt}}
\newcommand{\subappendix}[1] {\vspace{12pt}
        \refstepcounter{subappendixc}
        \noindent{\bf Appendix \thesubappendixc. {\kern1pt \bfit #1}}
	\par\vspace{5pt}}
\newcommand{\subsubappendix}[1] {\vspace{12pt}
        \refstepcounter{subsubappendixc}
        \noindent{\rm Appendix \thesubsubappendixc. {\kern1pt \tenit #1}}
	\par\vspace{5pt}}

\newcommand{\textlineskip}{\baselineskip=13pt}
\newcommand{\smalllineskip}{\baselineskip=10pt}

\def\eightcirc{
\begin{picture}(0,0)
\put(4.4,1.8){\circle{6.5}}
\end{picture}}
\def\eightcopyright{\eightcirc\kern2.7pt\hbox{\eightrm c}} 

\newcommand{\copyrightheading}[1]
	{\vspace*{-2.5cm}\smalllineskip{\flushleft
	{\footnotesize International Journal of Modern Physics B, #1}\\
	{\footnotesize $\eightcopyright$\, World Scientific Publishing
	 Company}\\
	 }}

\newcommand{\publisher}[2]{{\begin{center}\footnotesize\smalllineskip 
	Received #1\\
	Revised #2
	\end{center}
	}}

\def\abstracts#1#2#3{{
	\centering{\begin{minipage}{4.5in}\baselineskip=10pt\footnotesize
	\parindent=0pt #1\par 
	\parindent=15pt #2\par
	\parindent=15pt #3
	\end{minipage}}\par}} 

\def\keywords#1{{
	\centering{\begin{minipage}{4.5in}\baselineskip=10pt\footnotesize
	{\footnotesize\it Keywords}\/: #1
	\end{minipage}}\par}}

\renewenvironment{thebibliography}[1]			
	{\frenchspacing
	 \ninerm\baselineskip=11pt
	 \begin{list}{\arabic{enumi}.}
	{\usecounter{enumi}\setlength{\parsep}{0pt}
	 \setlength{\leftmargin 12.7pt}{\rightmargin 0pt} 
	 \setlength{\itemsep}{0pt} \settowidth
	{\labelwidth}{#1.}\sloppy}}{\end{list}}

\newcounter{itemlistc}
\newcounter{romanlistc}
\newcounter{alphlistc}
\newcounter{arabiclistc}

\newcommand{\fcaption}[1]{
        \refstepcounter{figure}
        \setbox\@tempboxa = \hbox{\footnotesize Fig.~\thefigure. #1}
        \ifdim \wd\@tempboxa > 5in
           {\begin{center}
        \parbox{5in}{\footnotesize\smalllineskip Fig.~\thefigure. #1}
            \end{center}}
        \else
             {\begin{center}
             {\footnotesize Fig.~\thefigure. #1}
              \end{center}}
        \fi}

\newcommand{\tcaption}[1]{
        \refstepcounter{table}
        \setbox\@tempboxa = \hbox{\footnotesize Table~\thetable. #1}
        \ifdim \wd\@tempboxa > 5in
           {\begin{center}
        \parbox{5in}{\footnotesize\smalllineskip Table~\thetable. #1}
            \end{center}}
        \else
             {\begin{center}
             {\footnotesize Table~\thetable. #1}
              \end{center}}
        \fi}

\def\@citex[#1]#2{\if@filesw\immediate\write\@auxout
	{\string\citation{#2}}\fi
\def\@citea{}\@cite{\@for\@citeb:=#2\do
	{\@citea\def\@citea{,}\@ifundefined
	{b@\@citeb}{{\bf ?}\@warning
	{Citation `\@citeb' on page \thepage \space undefined}}
	{\csname b@\@citeb\endcsname}}}{#1}}

\newif\if@cghi
\def\cite{\@cghitrue\@ifnextchar [{\@tempswatrue
	\@citex}{\@tempswafalse\@citex[]}}
\def\citelow{\@cghifalse\@ifnextchar [{\@tempswatrue
	\@citex}{\@tempswafalse\@citex[]}}
\def\@cite#1#2{{$\null^{#1}$\if@tempswa\typeout
	{IJCGA warning: optional citation argument 
	ignored: `#2'} \fi}}

\def\@citexo[#1]#2{\if@filesw\immediate\write\@auxout
        {\string\citation{#2}}\fi
\def\@citeao{}\@citeo{\@for\@citebo:=#2\do
        {\@citeao\def\@citeao{,}\@ifundefined
        {b@\@citebo}{{\bf ?}\@warning
        {Citation `\@citebo' on page \thepage \space undefined}}
        {\csname b@\@citebo\endcsname}}}{#1}}

\newif\if@cghi
\def\citeo{\@cghitrue\@ifnextchar [{\@tempswatrue
        \@citexo}{\@tempswafalse\@citexo[]}}
\def\citelowo{\@cghifalse\@ifnextchar [{\@tempswatrue
        \@citexo}{\@tempswafalse\@citexo[]}}
\def\@citeo#1#2{{{#1}\if@tempswa\typeout
        {IJCGA warning: optional citation argument
        ignored: `#2'} \fi}}

\def\pmb#1{\setbox0=\hbox{#1}
	\kern-.025em\copy0\kern-\wd0
	\kern.05em\copy0\kern-\wd0
	\kern-.025em\raise.0433em\box0}

\def\fnt#1#2{\footnotetext{\kern-.3em
	{$^{\mbox{\scriptsize #1}}$}{#2}}}

\def\fpage#1{\begingroup
\voffset=.3in
\thispagestyle{empty}\begin{table}[b]\centerline{\footnotesize #1}
	\end{table}\endgroup}

\def\runninghead#1#2{\pagestyle{myheadings}
\markboth{{\protect\footnotesize\it{\quad #1}}\hfill}
{\hfill{\protect\footnotesize\it{#2\quad}}}}
\font\tenrm=cmr10
\font\tenit=cmti10 
\font\tenbf=cmbx10
\font\bfit=cmbxti10 at 10pt
\font\ninerm=cmr9

\font\eightrm=cmr8

\def\qed{\hbox{${\vcenter{\vbox{			
   \hrule height 0.4pt\hbox{\vrule width 0.4pt height 6pt
   \kern5pt\vrule width 0.4pt}\hrule height 0.4pt}}}$}}

\renewcommand{\thefootnote}{\fnsymbol{footnote}}	

\def\bsc{{\sc a\kern-6.4pt\sc a\kern-6.4pt\sc a}}	
\def\bflatex{\bf L\kern-.30em\raise.3ex\hbox{\bsc}\kern-.14em 
T\kern-.1667em\lower.7ex\hbox{E}\kern-.125em X} 

\begin{document}

\runninghead{P. Fr{\"o}jdh, M. Howard \& K. B. Lauritsen}
{Directed Percolation and Systems with Absorbing States} 
\normalsize\textlineskip
\thispagestyle{empty}
\setcounter{page}{1761}

\copyrightheading{Vol.\ 15, No.\ 12 (2001) 1761---1797}

\vspace*{0.88truein}

\fpage{1761}
\centerline{\bf DIRECTED PERCOLATION AND}
\vspace*{0.035truein}
\centerline{\bf OTHER SYSTEMS WITH ABSORBING STATES:}
\vspace*{0.035truein}
\centerline{\bf IMPACT OF BOUNDARIES}

\vspace*{0.37truein}
\centerline{\footnotesize PER FR{\"O}JDH\footnote{email: frojdh@physto.se}}
\vspace*{0.015truein}
\centerline{\footnotesize\it Department of Physics, Stockholm University}
\baselineskip=10pt
\centerline{\footnotesize\it Box 6730, S-113 85 Stockholm, Sweden}
\vspace*{10pt}
\vspace*{10pt}
\centerline{\footnotesize MARTIN HOWARD\footnote{email: mjhoward@sfu.ca}}
\vspace*{0.015truein}
\centerline{\footnotesize\it Department of Physics, Simon Fraser University}
\baselineskip=10pt
\centerline{\footnotesize\it Burnaby, British Columbia, Canada V5A 1S6}
\vspace*{10pt}
\vspace*{10pt}
\centerline{\footnotesize KENT B{\AE}KGAARD LAURITSEN\footnote{email:
	 kbl@dmi.dk}}
\vspace*{0.015truein}
\centerline{\footnotesize\it Atmosphere Ionosphere Research Division}
\baselineskip=10pt
\centerline{\footnotesize\it Danish Meteorological Institute,
	2100 Copenhagen, Denmark}

\vspace*{0.21truein}

\publisher{(received date)}{(revised date)}

\vspace*{0.21truein}
\abstracts{
We review the critical behavior of nonequilibrium systems, such as
directed percolation (DP) and branching-annihilating random walks
(BARW), which possess phase transitions into absorbing states.
After reviewing the bulk scaling behavior of these models,
we devote the main part of this review to analyzing the impact of
walls on their critical behavior. We discuss the possible boundary 
universality classes for the DP and BARW models, which can be described
by a general scaling theory which allows for two 
independent surface exponents in addition to the bulk critical exponents. 
Above the upper critical dimension $d_c$, we review the use of mean
field theories, whereas in the regime $d<d_c$, where fluctuations are
important, we examine the application of field theoretic methods. Of
particular interest is the situation in $d=1$, which has been extensively
investigated using numerical simulations and series expansions.
Although DP and BARW fit into the same scaling theory,
they can still show very different surface behavior: for DP some
exponents are degenerate, a property not shared with the BARW model.
Moreover, a ``hidden'' duality symmetry of BARW in $d=1$
is broken by the boundary and this relates exponents and boundary
conditions in an intricate way.
}{}{}

\vspace*{10pt}

\newpage


\vspace*{1pt}\textlineskip	
\section{Introduction}	
\vspace*{-0.5pt}
\textheight=7.8truein
\setcounter{footnote}{0}
\renewcommand{\thefootnote}{\alph{footnote}}
\noindent
As is well known from the study of equilibrium statistical mechanics, 
boundaries have important effects on
systems close to criticality.\cite{binder,diehl,diehl2} Quantities
measured close to a wall can scale differently than in the bulk where,
for a given bulk universality class, various boundary universality
classes are possible. An analysis of the effects of boundaries
is clearly very important if one wishes to apply the theory of critical
phenomena to real physical systems. Although the theory of equilibrium 
surface critical phenomena is now well understood, the
equivalent theory for nonequilibrium systems is in a much less
advanced state. It is the purpose of this review to describe recent
progress in extending the theory of surface critical phenomena to
certain nonequilibrium problems.

The most prominent example of such a nonequilibrium dynamic system is
directed percolation (DP), which is the generic model for systems with
a continuous phase transition from an active to an absorbing state
from which the system cannot escape. DP describes the directed growth
of a cluster (i.e.\ growth in a preferred spatial direction or along
the time axis), where the growth rate is governed by a microscopic
growth probability $p$. For probabilities below a critical value,
$p < p_c$, the cluster will die after a finite time, which means
that the system gets trapped in the vacuum---the unique empty state.
On the other hand, for high enough growth probabilities
$p > p_c$, there is a finite probability that the cluster will always
remain active. Exactly at $p=p_c$, there is a
critical phase transition from the active into the absorbing 
state.\cite{kinzel} A whole range of systems possessing a phase
transition from a non-trivial active phase into a unique absorbing
state fall into this universality class. Some examples include
epidemics, chemical reactions, interface pinning/depinning,
the contact process, polynuclear growth, 
and certain cellular automata.\cite{dickman,hayereview} 

Although most models with absorbing states fall into the ubiquitous 
DP universality class, there are important exceptions. 
For instance, the model of {\it branching-annihilating 
random walks} with an even number of offspring (BARW) exhibits quite
different behavior\cite{tt,jensen:1994,cardy-tauber} and defines a
separate universality class. Other models in this class (at least in
one spatial dimension, $d=1$) include certain probabilistic
cellular automata,\cite{automata} monomer-dimer 
models,\cite{monomerdimer,hwang-etal:1998,inf_DP_DI} non-equilibrium
kinetic Ising models,\cite{menyhard-odor} and generalized DP with two
absorbing states (DP2).\cite{haye} These models escape from the DP
universality class by possessing an extra conservation law or
symmetry: for the BARW model, a ``parity'' conservation of the total
number of particles modulo $2$; for the other models, an underlying
symmetry between their absorbing states. Models with an infinite
number of absorbing states are believed to belong 
to the BARW class if the absorbing states can be arranged into two
symmetric groups (see Ref.\ \citeo{inf_DP_DI} and references therein).
On the other hand, if no higher symmetries between the absorbing states
are present, then such models will belong to the DP class.

In the present review we focus our attention on the impact of
walls on systems with nonequilibrium critical phase transitions into absorbing
states.\cite{janssen-etal,essam-etal:1996,lauritsen-etal,
dp-wall-edge,hayehari,ourprl,newjensen,menezes-moukarzel,
nancy,seattle,magnus,ourpre,grassberger-3d}
We will describe the boundary critical behavior of DP 
and BARW by using a variety of
methods, including mean field theory, scaling theory, field theory,
exact calculations, Monte-Carlo simulations, and series expansions.  
The rest of this review is organized as follows:
In Section~\ref{bulkmodels}
we introduce percolation and reaction-diffusion models
for DP and BARW for bulk systems (without boundaries).
Then in Section~\ref{bulkscaling} we quickly review the general bulk
scaling behavior of such systems. In Section~\ref{eqsurfcritbeh} we 
briefly discuss the surface critical behavior of
equilibrium systems, with a particular emphasis on the semi-infinite
Ising model. In Section~\ref{surfmodels} we then introduce walls into
our nonequilibrium models and discuss in detail some possible boundary
conditions in $d=1$. In
Sections~\ref{surfcritbeh}, \ref{scalingtheory}, \ref{sec:ft}, and
\ref{exres} 
we review the various theoretical techniques 
used to analyze the surface critical behavior of our nonequilibrium
models, including mean field, scaling, and field theories and also
exact calculations. In Section~\ref{sec:numres} we present results
for critical
exponents based on numerical work. In Section~\ref{otherwork} we
review recent work
in other directions concerning surfaces in DP-like systems.
Finally we give a summary and outlook in Section~\ref{summary}.

\section{Bulk DP and BARW}\label{bulkmodels}
\noindent
In this section we introduce the models for DP and BARW. 
We describe both reaction-diffusion versions of these models as
well as the probabilistic cellular automata frequently used in
simulations in $d=1$.

\subsection{DP}
\vglue 0pt
\noindent
The easiest way of introducing DP is to consider bond 
percolation on a directed lattice.
The update rules for bond DP in $d$ spatial dimensions
are then easily
defined: for each site at time $t$, form bonds with probability $p$
to the neighboring sites at time $t+1$.\cite{kinzel}

As is well known, various
reaction-diffusion models also fall into the DP universality
class.\cite{cardy-sugar,janssen1981,sundermeyer} The simplest of these
is defined by the following reaction scheme: 
\begin{eqnarray}
	& A \to A+A & {\rm ~with~rate~}\sigma \nonumber \\
	& A+A \to A & {\rm ~with~rate~}\lambda \label{dpreacs} \\
	& A \to \emptyset & {\rm ~with~rate~}\mu , \nonumber
\end{eqnarray} 
where the identical particles $A$ otherwise perform simple random walks
with diffusion constant $D$. 

\begin{figure}[htbp]
\vspace*{13pt}
\begin{center}
\leavevmode
\vbox{
\epsfxsize=3in
\epsffile{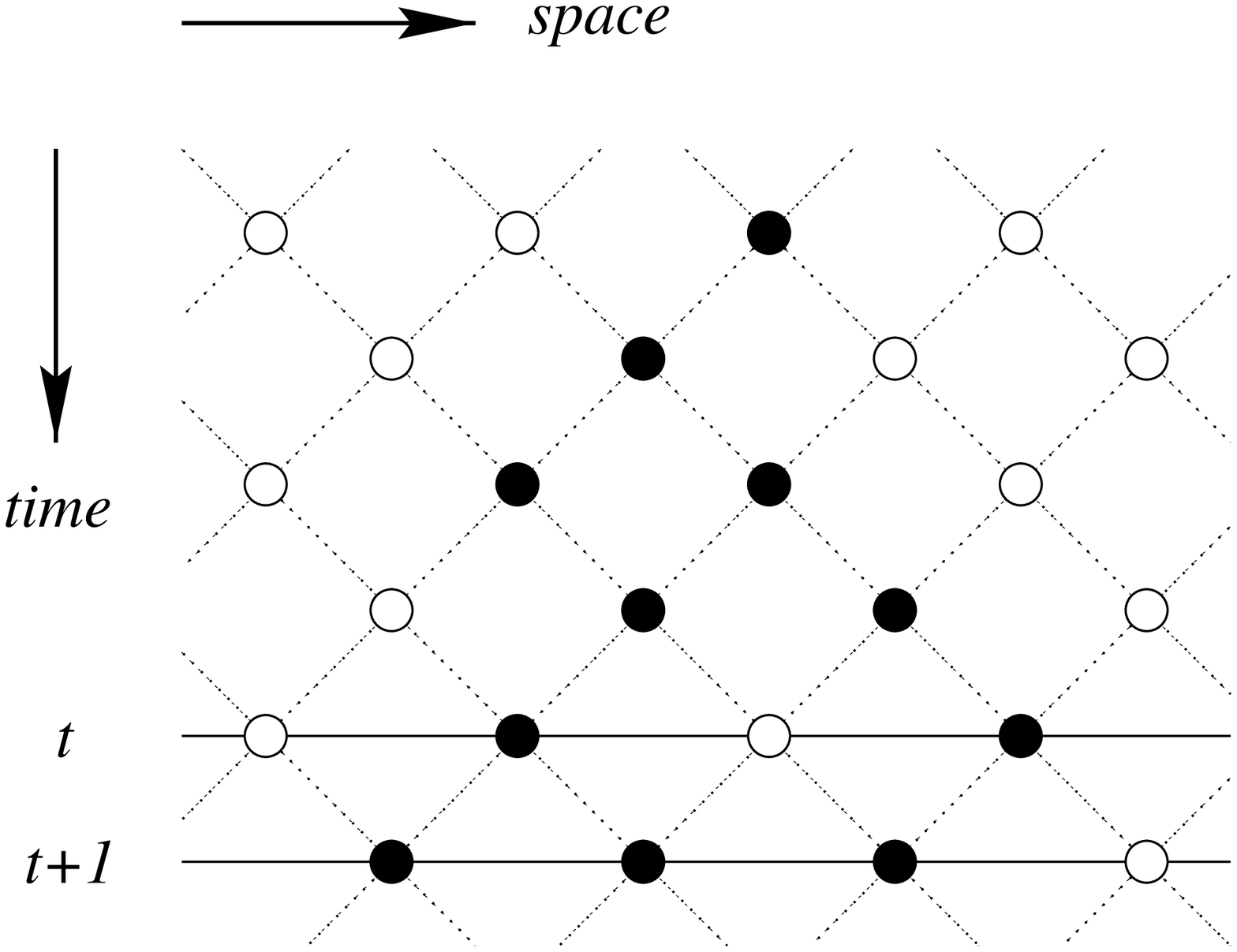}}
\end{center}
\vspace*{13pt}
\fcaption{Directed Percolation in terms of the Domany-Kinzel model,
where time flows vertically downwards. Black sites are
active ($A$) and white ones inactive ($I$). The state of each site at
time $t+1$ depends on the states of the neighboring sites at time $t$.}
\label{DPlattice}
\end{figure}

\begin{figure}[htbp]
\vspace*{13pt}
\begin{center}
\leavevmode
\vbox{
\epsfxsize=2.5in
\epsffile{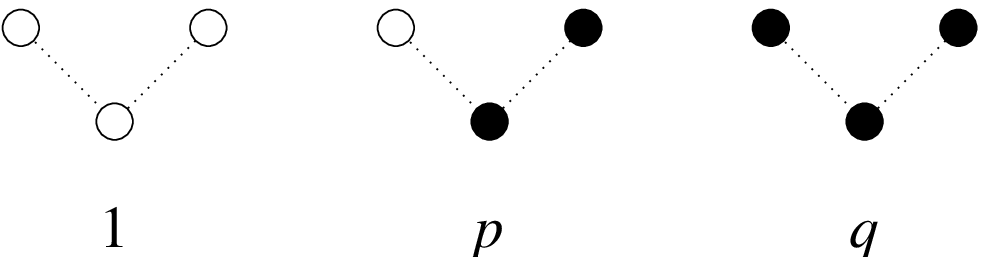}}
\end{center}
\vspace*{13pt}
\fcaption{Update probabilities for DP in terms of the parameters
$0 \leq p, q \leq 1$, where we have $q=p(2-p)$ for bond DP
and $q=p$ for site DP, respectively. Probabilities for the other
configurations follow from left-right symmetry and from
$P(A \mid \ldots) + P(I \mid \ldots) = 1$.}
\label{DPrules}
\end{figure}

\begin{figure}[htbp]
\vspace*{13pt}
\centerline{\hbox{
\epsfxsize=1.5in
\epsfbox{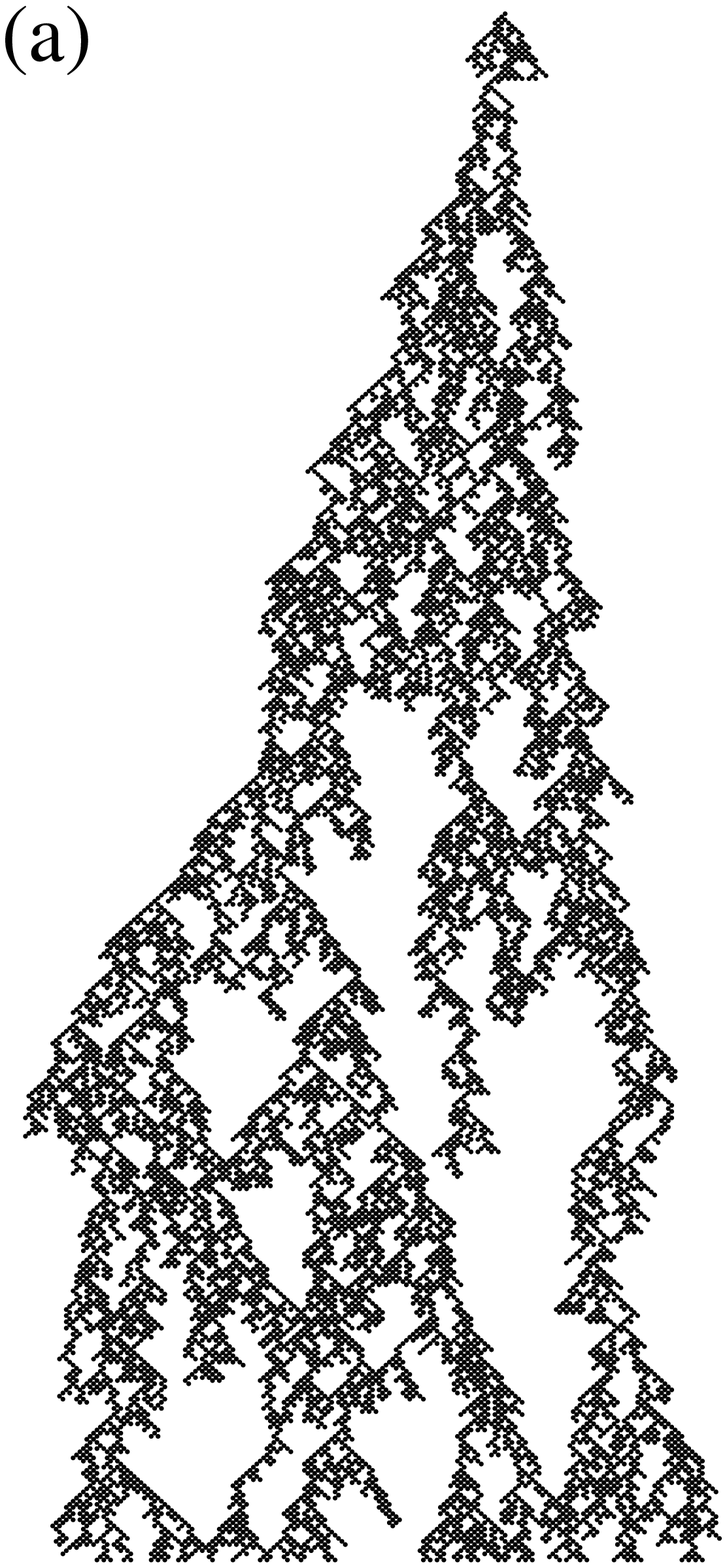}}
\epsfxsize=1.5in
\epsfbox{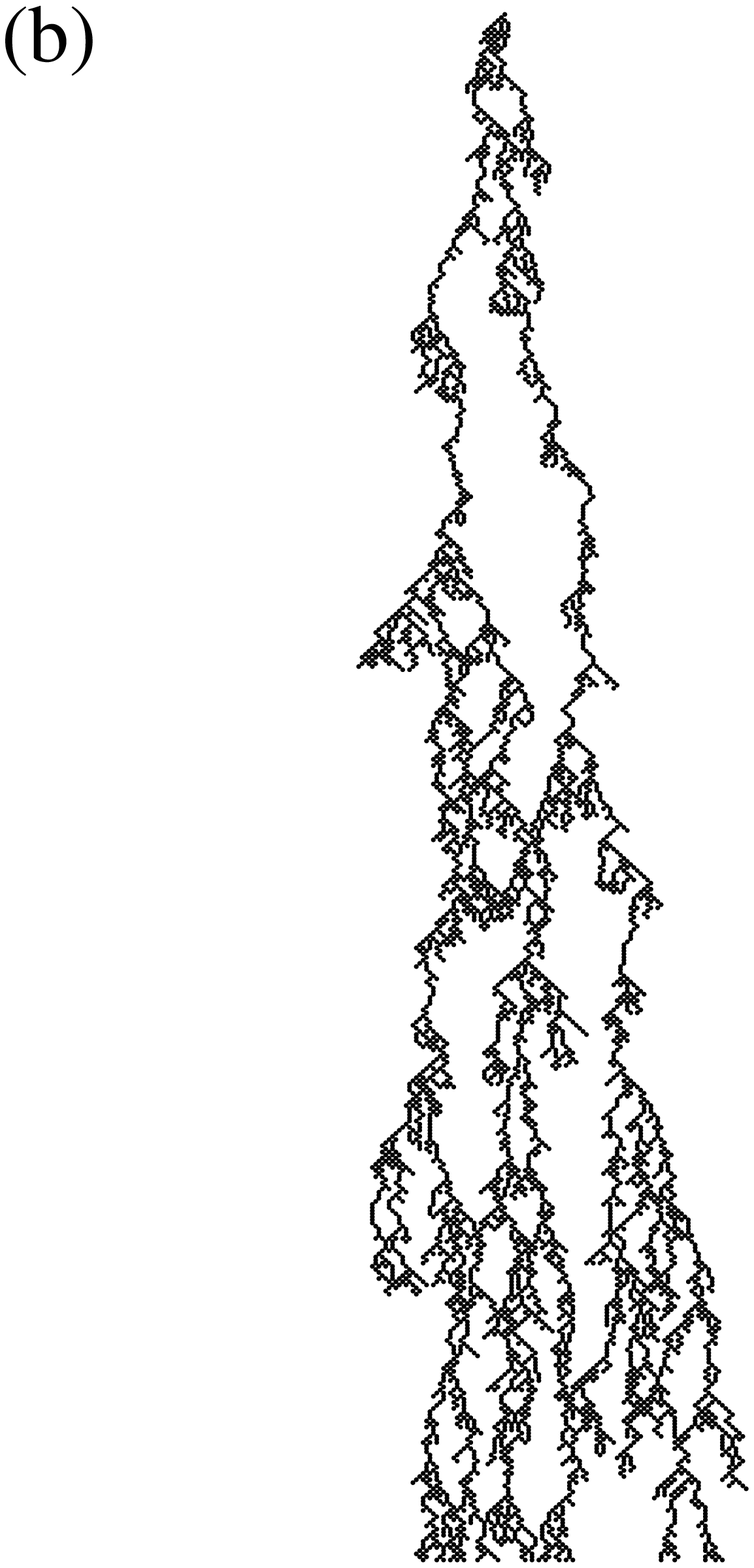}}
\vspace*{13pt}
\fcaption{(a) A DP cluster and (b) a DP2 cluster both grown from
a single seed in the bulk.} 
\label{dp2fig}
\end{figure}

In $d=1$, the DP universality class has also frequently been studied
by using the Domany-Kinzel model.\cite{domany1,domany2} In this
cellular automata model each site can either be active or inactive and
the probability for site $i$ to be updated to state $s_{i,t+1}$ at
time $t+1$ is given by an update probability $P(s_{i,t+1} | s_{i-1,t},
s_{i+1,t})$. See Figure~\ref{DPlattice} for a typical lattice
configuration and Figure~\ref{DPrules} for the update rules. An
example of a cluster grown from a single seed according to these rules
is shown in Figure~\ref{dp2fig}a.

\subsection{BARW}
\noindent

The BARW model is defined by a reaction-diffusion
system, with the following reaction
processes\cite{tt,jensen:1994,cardy-tauber} 
\begin{eqnarray}
	& A \to (m+1)A & {\rm ~with~rate~}\sigma_m \nonumber \\
	& A+A\to\emptyset & {\rm ~with~rate~}\lambda , \label{barwreacs}
\end{eqnarray}
where the identical particles $A$ otherwise perform simple random walks
with diffusion constant $D$.
For $m$ odd, the above model is known to belong to the DP universality
class, however for $m$ even we have what we refer to as the BARW 
universality class. Unless otherwise specified when we refer to 
the BARW model we will be referring to the even $m$ case.

\begin{figure}[htbp]
\vspace*{13pt}
\begin{center}
\leavevmode
\vbox{
\epsfxsize=3in
\epsffile{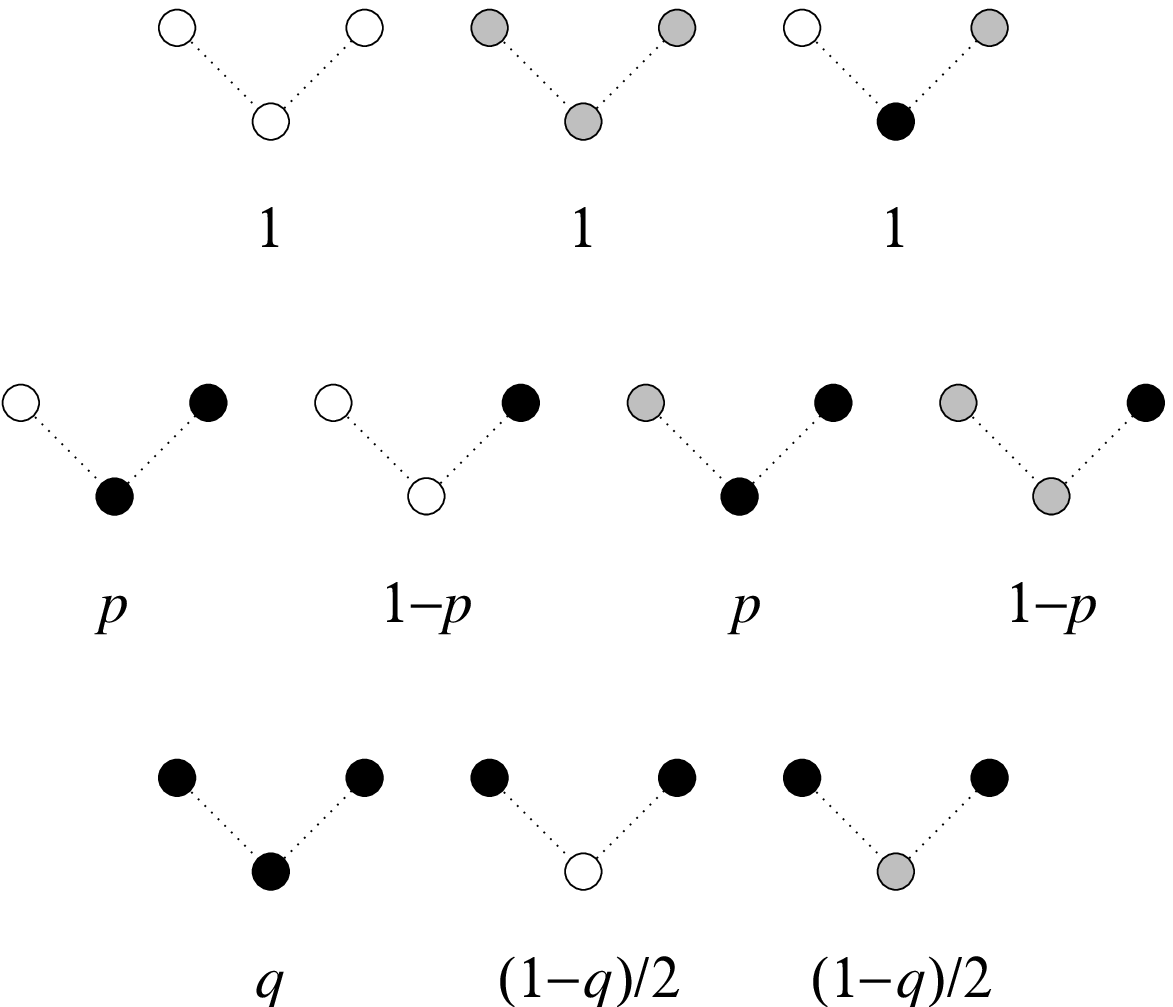}}
\end{center}
\vspace*{13pt}
\fcaption{Update probabilities for DP2: black sites are active ($A$),
whereas 
white and grey sites are in the inactive states $I_1$ and $I_2$,
respectively.  Probabilities for the other configurations follow from
left-right symmetry and from $P(A \mid \ldots) + P(I_1 \mid \ldots)
+ P(I_2 \mid \ldots) = 1$.}
\label{DP2rules}
\end{figure}

The BARW class has also been studied in $d=1$ using a generalized
Domany-Kinzel model (sometimes called DP$n$\cite{ourprl,ourpre})
introduced by Hinrichsen.\cite{haye} In this model each site can be
either active or in one of $n$ inactive states. For $n=1$ the update
rules are identical to those of the Domany-Kinzel model in
Figure~\ref{DPrules}, but for $n \geq 2$, the distinction between
regions of different inactive states is preserved by demanding that
they are separated by active ones. An example of a DP2 cluster is
shown in Figure~\ref{dp2fig}b. 
The DP2 model has two
symmetric absorbing states in which the system can become trapped. 
The update probabilities for $d=1$, where DP2 is known to
belong to the BARW universality class,\cite{haye} are given in
Figure~\ref{DP2rules}.

\break

\section{Bulk Scaling}\label{bulkscaling}
\begin{figure}[htbp]
\vspace*{13pt}
\begin{center}
\leavevmode
\vbox{
\epsfxsize=3in
\epsffile{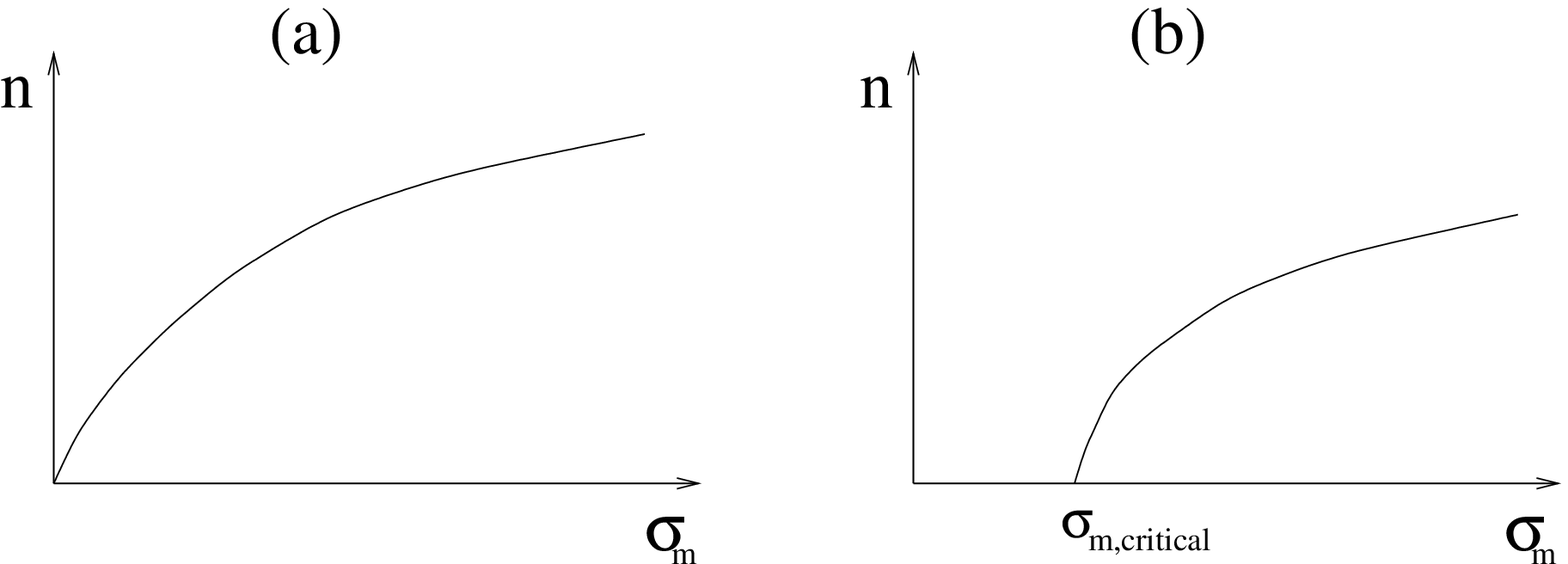}}
\end{center}
\vspace*{13pt}
\fcaption{Schematic bulk behavior for BARW of the density $n$ as a function of
the branching rate $\sigma_m$ for (a) $d\geq 2$ and (b) $d=1$.} 
\label{pbulk}
\end{figure}

\vglue 0pt
\noindent
The growth of both DP and BARW clusters in the bulk 
close to criticality can be summarized by a set of independent exponents. 
A natural choice is to consider $\nu_\perp$ and $\nu_\parallel$ 
which describe the divergence of the correlation lengths in space, 
\begin{equation}
	\xi_\perp \sim |\Delta|^{-\nu_\perp}, 
					\label{eq:xi_perp}
\end{equation}
and time 
\begin{equation}
	\xi_\parallel \sim |\Delta|^{-\nu_\parallel}.
					\label{eq:xi_parallel}
\end{equation}
Here the parameter $\Delta$
describes the deviation from the critical point. For DP and DP$n$, $\Delta 
= p_c - p$, whereas for the reaction-diffusion models in mean field 
theory $\Delta = \mu-\sigma$ for DP, and $\Delta=-m\sigma_m$ for BARW. 
We also need the order parameter exponent $\beta$, which can 
be defined in two {\it a priori} different ways: it is either governed by 
the percolation probability (the probability that a cluster grown from a 
finite seed never dies),
\begin{equation}
\label{P_bulk}
         P(t\to\infty,\Delta) \sim |\Delta|^{\beta_{\rm seed}}, \qquad
         \Delta < 0, 
\end{equation}
or by the coarse-grained density of active sites in the steady state,
\begin{equation}
\label{n(Delta)}
         n(\Delta) \sim |\Delta|^{\beta_{\rm dens}}, \qquad \Delta < 0.
\end{equation}
When $\Delta<0$ the system is said to be in an
{\it active} state, whereas for $\Delta=0$ the system is
{\it critical} (with an algebraically decaying density), and for
$\Delta>0$ (if applicable) the system is either {\it inactive} (DP)
or again critical (BARW).\cite{diffprl}
For the case of DP, it is known that $\beta$ is unique: 
$\beta_{\rm seed} = \beta_{\rm dens}$ in any dimension, both above and
below the upper critical dimension $d_c=4$.
This follows from time-reversal symmetry\cite{menezes-moukarzel}
and field theoretic 
considerations\cite{grassberger-torre,cardy-sugar} and has been 
verified by extensive numerical work.
The relation also holds for BARW in $d=1$,
a result first suggested by numerics and now backed up by an exact
duality mapping.\cite{schutz2} However, this exponent equality
is certainly not always true: if we consider the BARW mean field regime
valid for spatial dimensions $d>d_c=2$, then
the system is in a critical inactive state only for a zero branching
rate, where the density decays away as a power law.
However, any nonzero branching rate results in an active state, with
a nonzero steady state density\cite{cardy-tauber}
(see Figure~\ref{pbulk}a). As the branching rate is reduced towards zero, 
this density (\ref{n(Delta)}) approaches zero
continuously with the mean field exponent $\beta_{\rm dens}=1$. 
Nevertheless, for $d>2$, the survival probability 
(\ref{P_bulk}) of a particle cluster will be finite for {\it any\/} 
value of the branching rate, implying that 
$\beta_{\rm seed}=0$ in mean-field theory. This result follows from 
the non-recurrence of random walks in $d>2$. 

Field theoretically, DP is believed to be satisfactorily 
understood---the appropriate field theory\cite{cardy-sugar,sundermeyer} 
(sometimes called Reggeon Field 
Theory, see Section~\ref{Dpft}) is
well under control and the exponents have been computed to two loop
order in an $\epsilon=4-d$ expansion.\cite{janssen1981}
However, for the case of BARW (see Section~\ref{BARWft}), a
description of the $d=1$ case poses considerable difficulties for the field
theory.\cite{cardy-tauber} 
These stem from the presence of two critical dimensions: $d_c=2$ (above
which mean-field theory applies) and $d_c' \approx 4/3$. For $d>d_c'$
the behavior of Figure~\ref{pbulk}a holds, i.e.\ an active state
results for 
{\it any} nonzero value of the branching $\sigma_m$, whereas for
$d<d_c'$ the system is only active for $\sigma_m>\sigma_{m,{\rm
critical}}$,\cite{cardy-tauber} as shown in Figure~\ref{pbulk}b.
This means that the physical spatial dimension $d=1$ cannot be
accessed using controlled perturbative expansions down from the upper
critical dimension  $d_c=2$. Furthermore, for the
$\sigma_m<\sigma_{m,{\rm critical}}$ region, the system is {\it not}
inactive (in the sense of an exponentially decaying density). Instead
this entire phase is controlled by the annihilation fixed point of the
$A+A\to\emptyset$ process, where the density decays away as a power
law. Hence this phase should rather be considered as still being
critical. 

Despite the problems associated with BARW for $d<d_c'$,
we can still put forward a general scaling theory for DP and
BARW, valid both above and below their critical dimensions. However,
we must retain a possible distinction between $\beta_{\rm seed}$ and
$\beta_{\rm dens}$. For example, 
the average lifetime $\langle t\rangle$ of finite clusters can be
derived from the scaling form for the survival probability
\begin{equation}
	\label{survbulk}
	P(t,\Delta)=|\Delta|^{\beta_{\rm seed}}\varphi(t/\xi_{\parallel}) .
\end{equation} 
We then find 
\begin{equation}
	\langle t \rangle = \int t \, P(t,\Delta) ~ dt \sim
	|\Delta|^{-\tau}, 
					\label{eq:<t>-bulk}
\end{equation}
where 
\begin{equation}
	\tau = \nu_\parallel - \beta_{\rm seed}. 
					\label{eq:tau}
\end{equation}
The appropriate scaling
form for the density $n({\bf x},t)$, given that the cluster was started
at ${\bf x}={\bf 0},t=0$, is 
\begin{equation}
	\label{densbulk}
	n(x,t,\Delta)=|\Delta|^{\beta_{\rm seed}+\beta_{\rm
	dens}}f(x/\xi_{\perp},t/\xi_{\parallel}) .
\end{equation}
Notice that rotational symmetry about the seeding point ${\bf x}=0$
implies that
the spatial coordinates enter the scaling function only as
$x=|{\bf x}|$, the distance from the seeding point.
Using the expression (\ref{densbulk}) we see that the average mass of
finite clusters scales as
\begin{equation}
	\langle s \rangle = \int n(x,t,\Delta) ~ d^d x \, dt
			\sim |\Delta|^{-\gamma}
					\label{eq:<s>}
\end{equation}
where $\gamma$ is related to the other
exponents via the following hyperscaling relation:
\begin{equation}
        \label{gen_bulk_hyperscaling}
        \nu_{\parallel} +d\nu_{\perp} 
                = \beta_{\rm seed} + \beta_{\rm dens} +\gamma .
\end{equation} 
Note that Eq.~(\ref{gen_bulk_hyperscaling}) is consistent with the
distinct upper critical dimensions for BARW and DP. Using the above
mean field values for BARW and $\nu_\perp=1/2$, $\nu_\parallel=1$, and 
$\gamma=1$, we verify $d_c = 2$. 
In contrast, for DP one has the mean-field exponents $\beta_{\rm
dens}=\beta_{\rm seed} = 1$, $\gamma=1$, and $d_c = 4$.
In Tables \ref{table-DP-exponents} and \ref{table-DP2-exponents}
we list the exponents for DP
(see Refs.\
\citeo{essam-etal:1996,jensen:1996,1d,1d-2,
grassberger-2d,grassberger-zhang,iwan-3d,briefreport,jensen:1999} and
references therein)
and BARW 
(see Refs.\ \citeo{haye,jensen:1997,zhong:1995} and references therein).

\begin{table}[htbp]
\tcaption{Critical exponents for DP. 
  }
\centerline{\footnotesize\smalllineskip
\begin{tabular}{l|l|l|l|c}\\
              & \makebox[20mm]{$d=1$} & \makebox[16mm]{$d=2$} 
              & \makebox[16mm]{$d=3$} & \makebox[16mm]{Mean Field} \\
\hline
$\beta_{\rm dens}=\beta_{\rm seed}$  
                       & ~0.276 486(8)  & ~0.583(4)  & ~0.805(10) &   1  \\
$\nu_\parallel$        & ~1.733 847(6)  & ~1.295(6)  & ~1.105(5)  &   1  \\
$\nu_\perp$            & ~1.096 854(4)  & ~0.733(4)  & ~0.581(5)  &  1/2 \\
$\tau$                 & ~1.457 362(14) & ~0.711(7)  & ~0.298(12) &   0  \\
$\gamma$               & ~2.277 730(5)  & ~1.593(7)  & ~1.232(12) &   1  \\
\end{tabular}}
\label{table-DP-exponents}
\end{table}

\begin{table}[htbp]
\tcaption{Critical exponents for BARW.
  }
\centerline{\footnotesize\smalllineskip
\begin{tabular}{l|l|c}\\
              & \makebox[16mm]{$d=1$} & \makebox[16mm]{Mean Field} \\
\hline
$\beta_{\rm dens}$     & ~0.922(5)    &   1  \\ 
$\beta_{\rm seed}$     & ~0.93(5)     &   0  \\ 
$\nu_\parallel$        & ~3.22(3)     &   1  \\ 
$\nu_\perp$            & ~1.84(2)     &  1/2 \\ 
$\tau$                 & ~2.30(3)     &   1  \\  
$\gamma$               & ~3.22(3)     &   1  \\   
\end{tabular}}
\label{table-DP2-exponents}
\end{table}

\section{Equilibrium Surface Critical Behavior}
\label{eqsurfcritbeh}
\noindent
We will begin our discussion of boundaries in critical systems by
briefly reviewing the theory of surface critical
phenomena for equilibrium systems,\cite{diehl2}  
with particular emphasis on the Ising
model. Defined on a half space, the semi-infinite Ising model serves
as the canonical example of a system exhibiting surface critical
behavior.  

Each bulk Ising spin has the same number $2d$ of nearest neighbors.
However, by cutting the lattice in half, spins next to the wall
possess fewer neighbors than those in the bulk and are therefore
more weakly coupled. At high temperatures $T$, the system is
disordered and thus correlations are short ranged. Hence
the effects of the wall are localized to a thin layer along the
wall and decay away exponentially into the bulk with a length
scale governed by the bulk correlation length. However, as the
temperature 
is lowered the system will become critical at $T=T_c$, the critical
temperature of the infinite system. At this point, where the
correlation length diverges, the decay of the perturbation introduced
by the wall is algebraic, and hence the impact of the wall on the
scaling behavior is very marked. The two-point spin-spin correlation
function decays with a new exponent along the wall, independent of the
two exponents required to describe bulk Ising criticality. In this
case, the critical behavior of the surface is referred to as the {\em
ordinary transition}. In two dimensions, the boundary critical
exponent 
governing the decay of the two point correlation function is known
exactly.\cite{mccoy-wu} 

We could also imagine replacing all the spin couplings at the wall by
stronger ones than in the bulk. Starting from the disordered state and 
lowering $T$, we will now have a situation where the strongly coupled
wall spins may potentially undergo a phase transition already at
$T>T_c$. In this case, the wall spins order  
independently of the bulk which remains disordered at this
temperature. Of course, if the boundary is of sufficiently low
dimension (e.g.\ the one dimensional boundary of a semi-infinite $2d$
Ising model) then such ordering cannot occur at finite
temperatures. But, for $d=3$, the
surface is two dimensional and in these circumstances can undergo a 
so called {\em surface transition}. This surface transition (if it
exists) lies in the same universality class as a $d-1$
dimensional bulk transition. By further reducing the temperature,
the bulk will then order at $T=T_c$ in the presence of an already 
ordered surface. In this case, the critical behavior of the boundary
is referred to as the {\em extraordinary
transition}. This is quite different to the ordinary transition
discussed above, where the surface was compelled to order through
its coupling to the ordered bulk. At the
extraordinary transition, where $T=T_c$, the
system is again critical, but the correlations close to  
the wall differ from the case of the ordinary transition, and are
governed by another independent boundary critical exponent. 

In principle we may tune the wall couplings such that the boundary
transition takes place at $T=T_c$. Another name for this point where
the bulk and boundary transitions coincide is the {\em special
transition}. It is a multicritical point and connects the lines of
ordinary, extraordinary and surface critical points in the phase
diagram of bulk and wall couplings. 

For $d=2$ the surface and special transitions cannot take place, since 
the surface is one-dimensional. However, an alternative way of
ordering the surface layer is to introduce a magnetic field that only
couples to wall spins. The boundary phase transition at $T=T_c$ in the
presence 
of such a field is called a {\em normal transition}. It turns out that
the exact mechanism responsible for ordering the wall is unimportant,
and thus the normal transition is in fact equivalent to the
extraordinary transition.\cite{normal-extraord}
Hence, in two dimensions, we only need to distinguish between two
universality classes. These can be accessed using the following
boundary conditions: a free (open) boundary will give an ordinary
transition and a 
fixed boundary will give an extraordinary transition. For both cases,
the corresponding exponents are exactly known. Even the four-point 
functions governing the universal cross-over from wall to bulk 
correlations are exactly known from conformal field
theory.\cite{cardy-cft} 

Another interesting aspect of the Ising model is the Kramers-Wannier
duality, which is a mapping of the Ising model from 
disorder to order and vice versa.\cite{kramers-wannier}
The self dual point defines $T_c$. However, by introducing a surface,  
the self-duality at $T=T_c$ is broken
and it is straightforward to show that under the
duality open boundaries are mapped to fixed boundaries and vice
versa. Hence, the duality now maps the two
boundary universality classes onto one another. This  
observation is useful to bear in mind when analyzing the BARW model,
where we will see that a self-duality broken by the wall again
relates boundary universality classes in a subtle way. 

\section{DP and BARW with Walls}
\label{surfmodels}
\noindent

We now turn our attention to an understanding of the boundary critical   
properties of the nonequilibrium DP and BARW models. We will begin by
detailing the boundary conditions of these models.

For the reaction-diffusion version of DP, the modification is simple,
we simply allow for the DP reactions (\ref{dpreacs}) to occur at (potentially)
different rates on the surface as compared to their values in the
bulk. By tuning these boundary reaction rates, we can access the
various boundary universality classes, in a similar way as was achieved
in the Ising model by tuning the surface spin couplings.

For the case of BARW, the situation is a little more complicated.
The basic idea is that on the surface we may include not
only the usual branching and annihilation reactions (\ref{barwreacs})
but potentially also a parity symmetry breaking $A\to\emptyset$
reaction. Depending on whether or not the $A\to\emptyset$ reaction is
actually present, we may then expect different boundary universality
classes according to whether the symmetry of the bulk is broken or
respected at the surface. We note that a somewhat similar situation in
an equilibrium system has recently been analyzed in
Ref.\ \citeo{drewitz}. 

\begin{figure}[bthp]
\vspace*{13pt}
\begin{center}
\leavevmode
\vbox{
\epsfxsize=3in
\epsffile{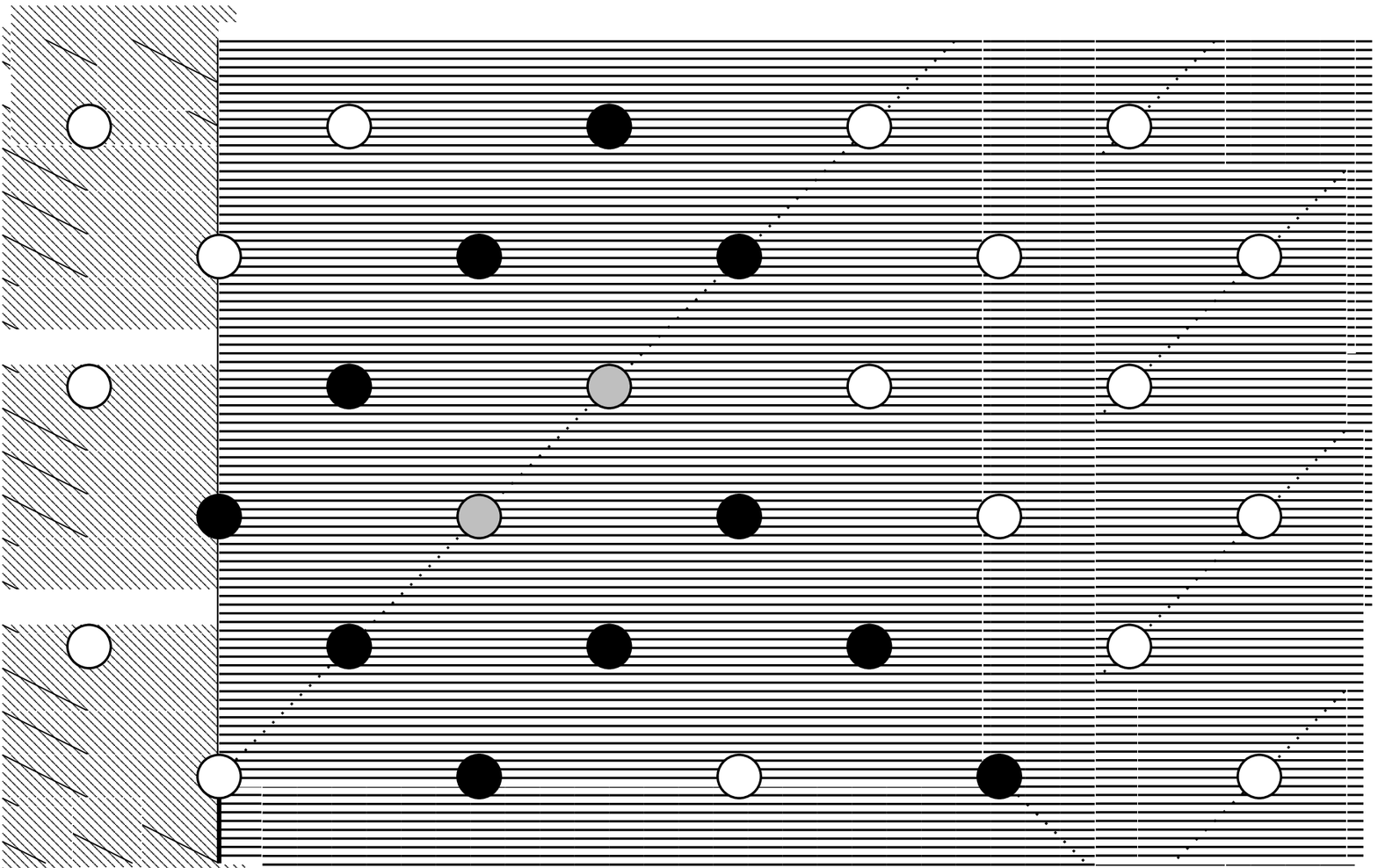}}
\end{center}
\vspace*{10pt}
\fcaption{DP2 with an inactive boundary condition (IBC).}
\label{DP2_IBC}
\end{figure}

For the case of the $d=1$ cellular automata introduced in 
Section~\ref{bulkmodels}, 
there are several obvious boundary conditions. The simplest 
is just to cut off the lattice. This is equivalent to
introducing boundary sites which are forced to be in one of the
inactive states. We will refer to this case as the inactive
boundary condition (IBC), and, for DP$2$, we choose
inactivity of type $1$ to the left of the boundary, 
see Figure~\ref{DP2_IBC}. Apart from imposing the state of these sites
within the wall, the sites at the wall and those in the bulk are
updated by the rules in Figures~\ref{DPrules} and~\ref{DP2rules}.

\begin{figure}[bhtp]
\vspace*{13pt}
\begin{center}
\leavevmode
\vbox{
\epsfxsize=3in
\epsffile{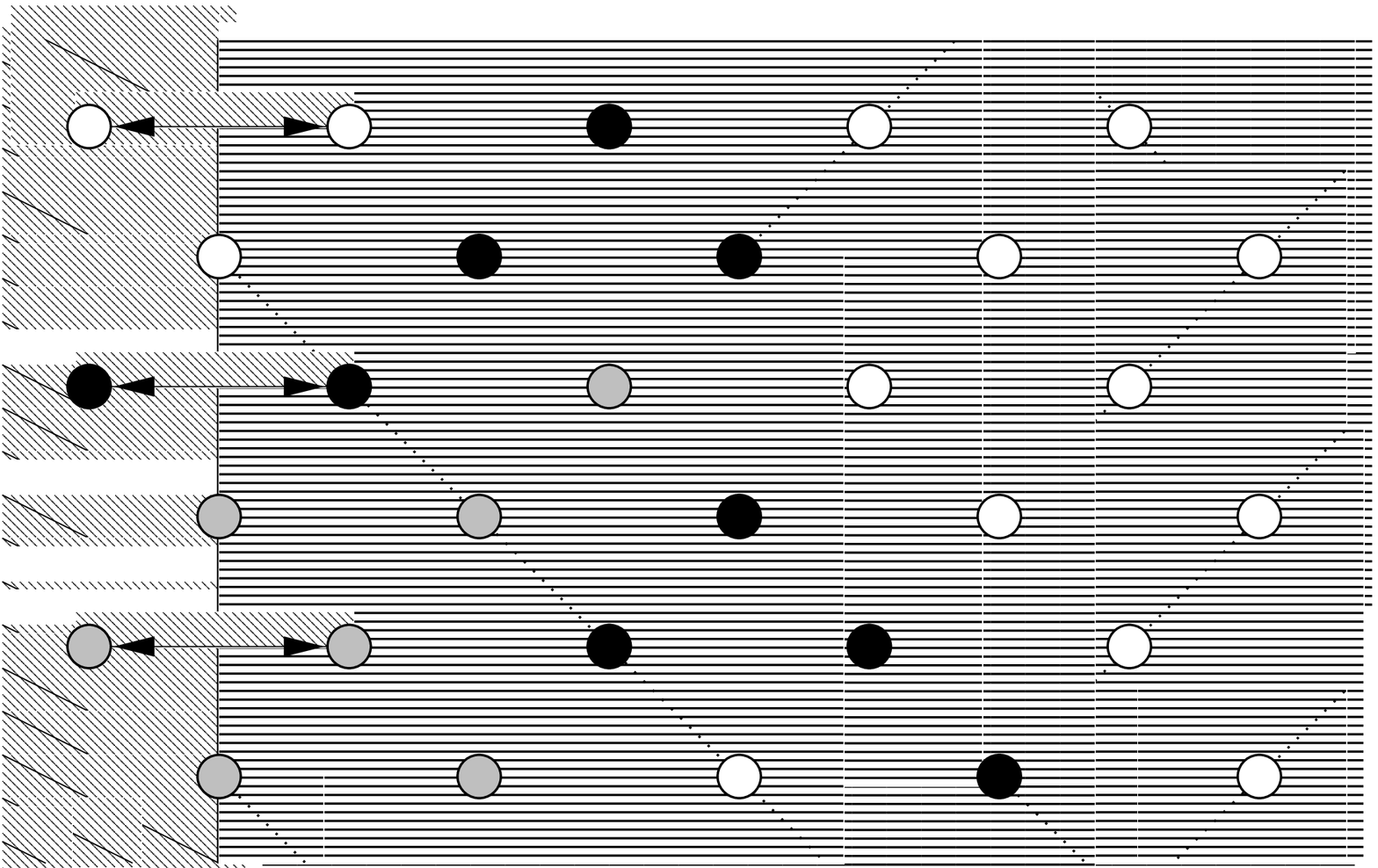}}
\end{center}
\vspace*{10pt}
\fcaption{DP2 with a reflecting boundary condition (RBC).}
\label{DP2_RBC}
\end{figure}

Next we consider the reflecting boundary condition (RBC) where the
wall acts like a mirror so that the sites within the wall
are always a mirror image of those next to the wall, see
Figure~\ref{DP2_RBC}. For DP2, one can see that there is a qualitative
difference between the IBC and the RBC. For the latter, regions of
type-$2$ inactivity can get trapped at the wall and the only way for
these regions to disappear is to wait for the cluster to return.
On the other hand, for the IBC, such regions are never trapped. 

\begin{figure}[tbhp]
\vspace*{13pt}
\begin{center}
\leavevmode
\vbox{
\epsfxsize=3in
\epsffile{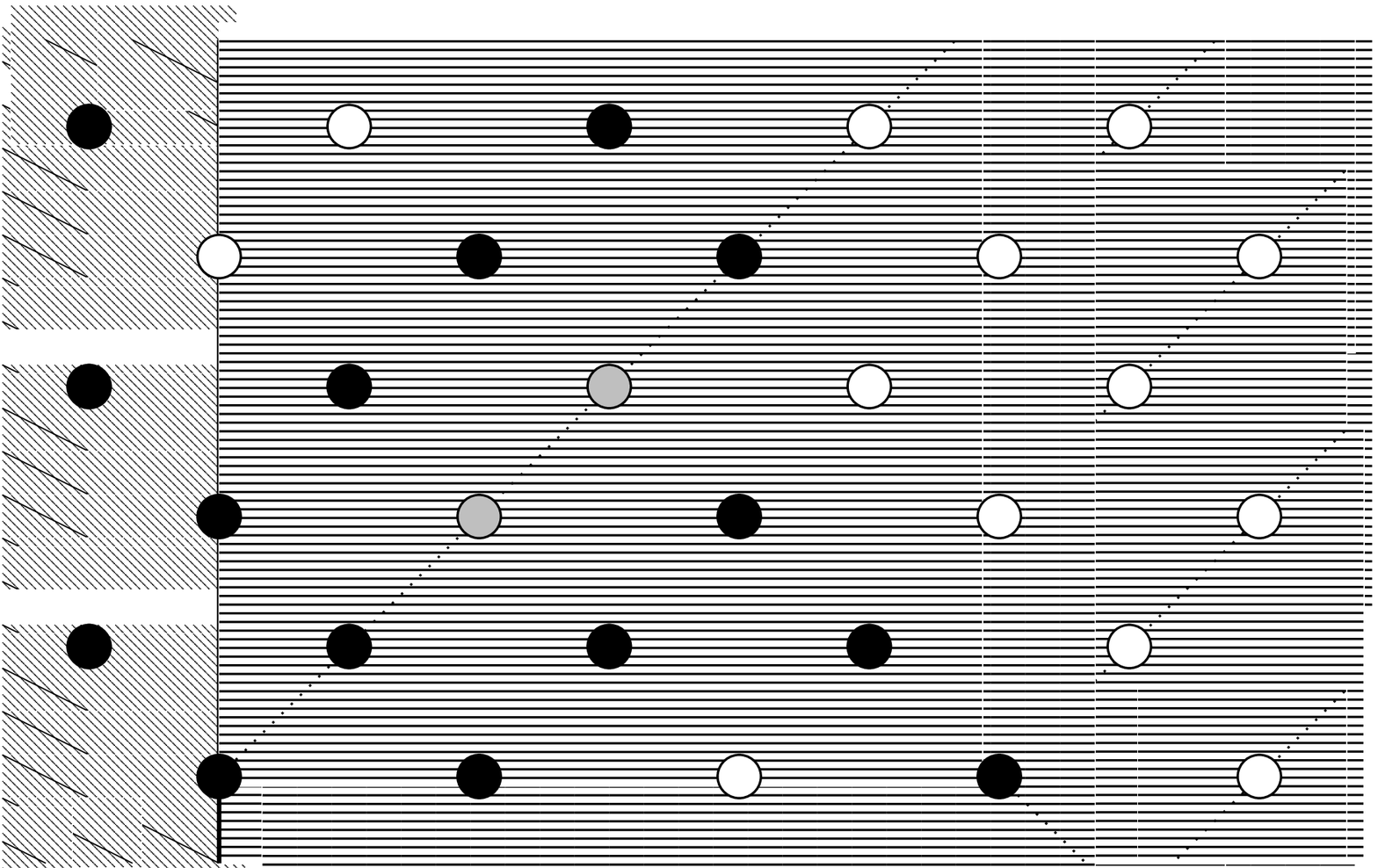}}
\end{center}
\vspace*{13pt}
\fcaption{DP2 with an active boundary condition (ABC).}
\label{DP2_ABC}
\end{figure}

A third type of boundary condition is the active boundary condition
(ABC) where the sites within the wall are forced to be active, see
Figure~\ref{DP2_ABC}. 
In this case the cluster will never die completely as the wall will
always be active and can always induce new clusters. 

Although it is not readily apparent, it turns out that these three
boundary conditions exhaust all possible boundary universality classes
for the BARW model in $d=1$. We will return later in
Sections~\ref{dpira} and \ref{barwira} to the question of which
boundary universality class is to be associated with each of the
IBC, RBC, and ABC boundary conditions.

\break

\section{Boundary Critical Behavior of DP and BARW}
\label{surfcritbeh}
\noindent

\subsection{Surface DP}
\label{sdpgendisc}

\begin{figure}[bthp]
\vspace*{13pt}
\centerline{\hbox{
\epsfxsize=1.5in
\epsfbox{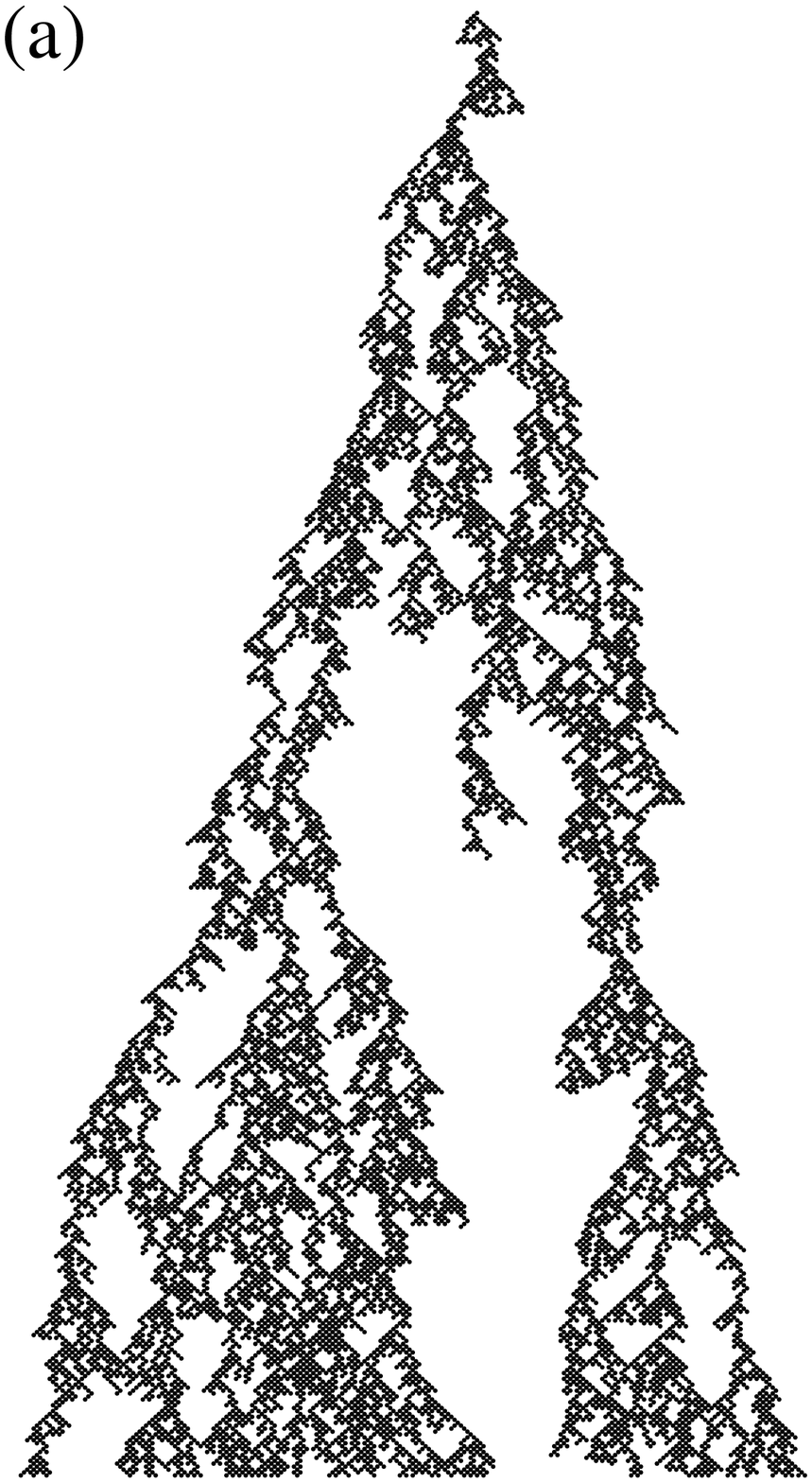}}
\epsfxsize=1.5in
\epsfbox{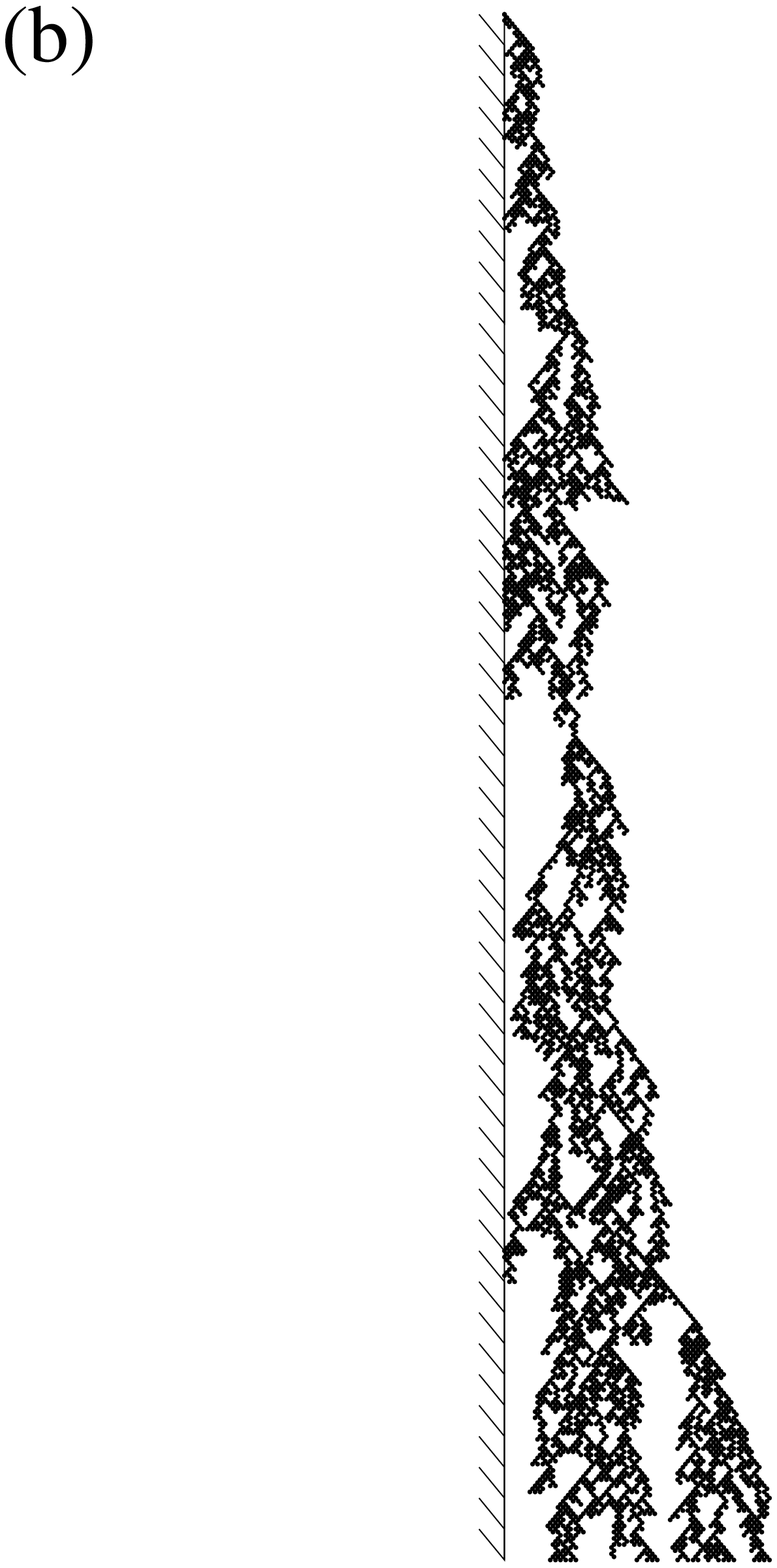}}
\vspace*{13pt}
\fcaption{DP clusters in the Domany-Kinzel model grown from a single
seed (a) in the bulk and (b) next to an impenetrable wall.} 
\label{dpfig}
\end{figure}

\noindent
First of all, let us examine the effects of introducing a $d-1$ dimensional 
wall at $x_\perp=0$ [${\bf x}=({\bf x_{\parallel}},x_{\perp})$] into a
DP process. Note that the labels parallel ($\parallel$) and perpendicular 
($\perp$) refer here to directions relative to the wall (and not 
relative to the time direction). An example of
such a cluster grown close to a wall is shown in Figure~\ref{dpfig}b.

\break

\begin{figure}[tbhp]
\begin{center}
\leavevmode
\vbox{
\epsfxsize=2.5in
\epsffile{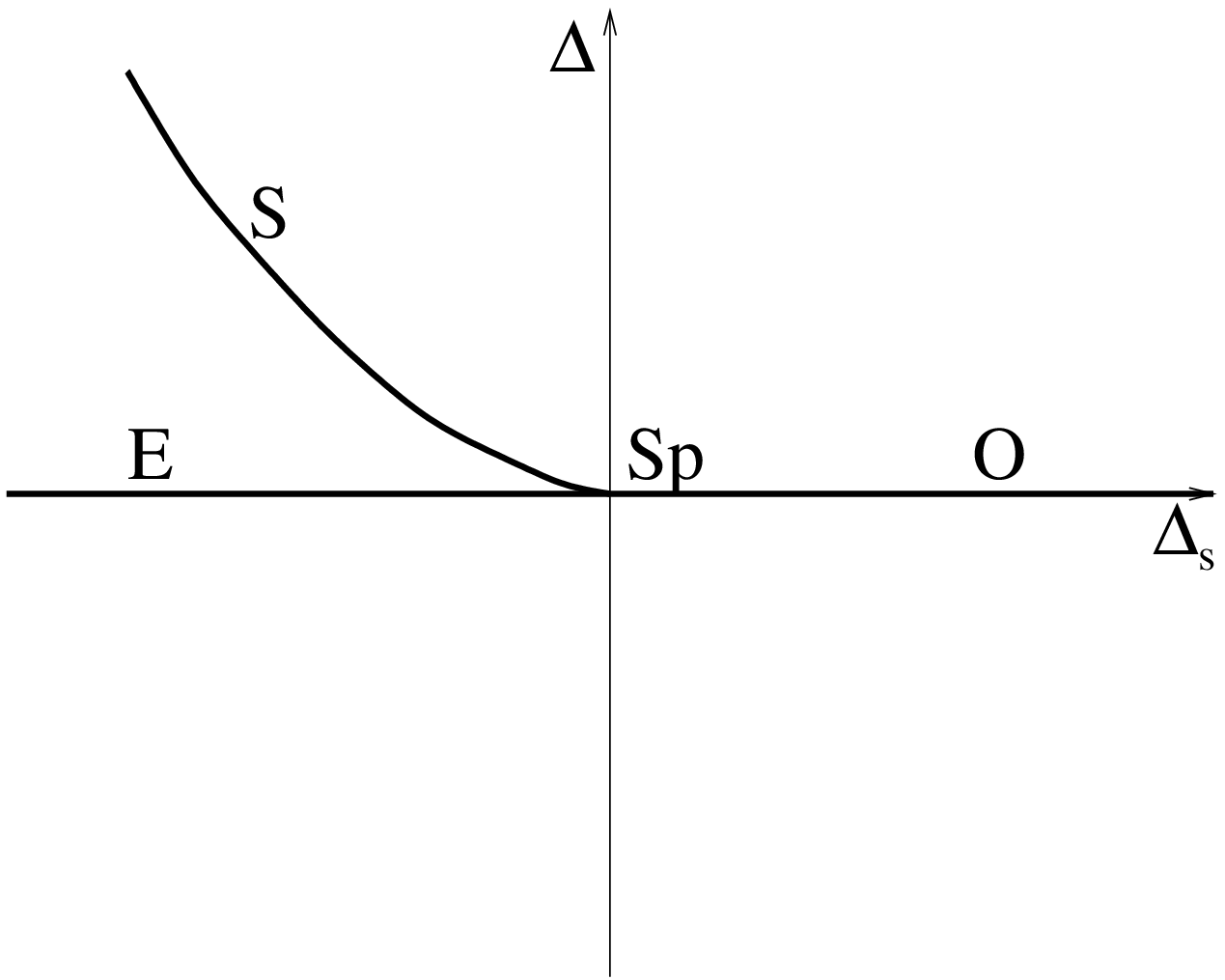}}
\end{center}
\vspace*{6mm}
\fcaption{Schematic mean field phase diagram for boundary DP. The
transitions are labeled by O=ordinary, E=extraordinary, S=surface, and 
Sp=special.} 
\label{dpps}
\end{figure}

Following on from our discussion of boundaries in equilibrium systems
(see Section~\ref{eqsurfcritbeh}), it is not difficult to justify a
schematic phase diagram for boundary DP (see Figure~\ref{dpps} and also
Ref.\ \citeo{janssen-etal}). In Figure~\ref{dpps}, $\Delta$ and
$\Delta_s$ represent, respectively, the deviations of the bulk and
surface from criticality. For $\Delta_s>0$ and as $\Delta\to 0$, we
see an ordinary transition, since in that case we expect the bulk to
order in a situation where the boundary, if isolated, would be
disordered. On the other hand, for $\Delta>0$ and for $\Delta_s$
sufficiently negative, we expect the boundary to order even while the
bulk is disordered, i.e.\ the surface transition. Then for
$\Delta_s<0$ and $\Delta\to 0$, the bulk will order in the presence of
an already ordered boundary, i.e.\ an extraordinary transition for the
boundary. Finally at $\Delta=\Delta_s=0$, where all the critical lines
meet, and where both the bulk and isolated surface would be critical,
we expect a multicritical point, i.e.\ the special transition.

The bulk exponents are, of course, unchanged by the presence of a
surface and, furthermore, one can show that the correlation
length exponents on the boundary are also the same as
in the bulk.\cite{diehl,janssen-etal} 
In the following, for conciseness, we will concentrate solely on the
ordinary transition; more details on some of the other transitions can
be found in Refs.\ \citeo{janssen-etal,ourpre}. At the ordinary
transition, one finds just {\it one} extra independent exponent associated with
the boundary: this can be taken to be the surface density exponent
$\beta_{1,{\rm dens}}$. This is defined from the steady-state density
at the wall, where we have
\begin{equation}
\label{ndpsc}
	n(x_\perp=0,\Delta) \sim |\Delta|^{\beta^{\rm O}_{1,{\rm dens}}},
	                    \qquad  \Delta<0. 
\end{equation}
In principle one could also allow for a second
type of surface $\beta_1$ exponent, one defined from a
survival probability for clusters started on the wall
\begin{equation}
\label{pdpsc}
	P_1(t\to\infty,\Delta) \sim |\Delta|^{\beta^{\rm O}_{1,{\rm seed}}},
                               \qquad \Delta<0.
\end{equation}
However, the surface exponents here show a similar pattern to their
bulk counterparts and fulfill
$\beta^{\rm O}_{1, \rm seed} = \beta^{\rm O}_{1,{\rm dens}}
=\beta_1^{\rm O}$, 
as can be shown using a time-reversal symmetry argument (cf.\ Ref.\ 
\citeo{menezes-moukarzel}) or by a field-theoretic derivation of
an appropriate correlation function (see Section \ref{Dpft} and
also Ref.\ \citeo{dp-wall-edge}).  

\subsubsection{Mean field theory for surface DP}
\noindent
In this section, we very briefly review some simple results for
boundary DP at the mean field level, focusing again on the 
ordinary transition. A more general analysis can be found in
Ref.\ \citeo{ourpre} (see also Appendix C of Ref.\
\citeo{nancy}). The equation describing mean field DP with a
surface is 
\begin{equation}
	\label{dpmfteq}
	\partial_t n = D\nabla^2 n -\Delta n -\lambda n^2 ,
\end{equation}
with the boundary condition 
\begin{equation}
	D \partial_{x_{\perp}}n|_{x_{\perp}=0}= \Delta_s n|_{x_{\perp}=0}.
						\label{eq:BCgeneral}
\end{equation}
Here the variable $\Delta=\mu-\sigma$ is the difference
between the rates for the $A\to\emptyset$ and $A\to A+A$
processes. Similarly we have the surface variable $\Delta_s$,
and the bulk quadratic term is due to the reaction $A+A\to A$. 
{}From the above Eq.~(\ref{dpmfteq}), the bulk mean field exponents 
can easily be computed: $\nu_{\parallel}=1$, $\nu_{\perp}=1/2$, and
$\beta=1$. Furthermore, with the inclusion of a boundary, we see that
the correlation length exponents are unchanged at the wall but the
surface $\beta_1$ exponents are altered. If we are interested in the mean
field steady state, then we can replace Eq.~(\ref{dpmfteq}) with
\begin{equation}
	\label{dpmftred}
	D n'' -\Delta n -\lambda n^2 =0 ,
\end{equation}
where $n''\equiv d^2n/dx_{\perp}^2$. The appropriate boundary
condition~(\ref{eq:BCgeneral}) is given by 
\begin{equation}
	D n'_s = \Delta_s n_s, 
                              \label{eq:BC}
\end{equation}
where $n_s=n|_{x_{\perp}=0}$, and
$n'_s=dn/dx_{\perp}|_{x_{\perp}=0}$. Multiplying Eq.~(\ref{dpmftred}) by
$n'$ and integrating, we have
\begin{equation}
\label{int1}
{1\over 2}Dn'^2-{1\over 2}\Delta n^2-{1\over 3}\lambda n^3 +C =0 ,
\end{equation}
where $C$ is a constant of integration. Using the bulk results $n'=0$,
and $n=(-\Delta)/\lambda$ for $\Delta<0$, or $n=0$ for $\Delta>0$,
we have 
\begin{eqnarray}
\label{int2a}
& {\Delta_s n_s\over D} = -\left[{\lambda\over D}\right]^{1/2}\left(n_s-
{|\Delta|\over\lambda}\right)\left({2\over 3}n_s+{|\Delta|\over
3\lambda}\right)^{1/2} & [\Delta<0]
\hspace{-.1in} \\
\label{int2b}
& {\Delta_s n_s\over D} = -\left[{\lambda\over
D}\right]^{1/2}n_s\left({2\over 
3}n_s+{\Delta\over\lambda}\right)^{1/2} & [\Delta>0]
\hspace{-.1in} 
\end{eqnarray}
where we have also used the boundary condition~(\ref{eq:BC}).
Considering the ordinary transition where $\Delta_s>0$ and
$\Delta\to 0^-$, we expect $n=|\Delta|/\lambda\gg n_s$,
and thus Eq.~(\ref{int2a}) yields $n_s\propto|\Delta|^{3/2}$, giving
the mean field value $\beta_1^{\rm O}=3/2$.

\break

\subsection{Surface BARW}
\label{BARWsurface}
\vglue 0pt
\noindent
For the case of BARW, we expect a similar picture to hold as that
described for DP in Section~\ref{sdpgendisc}. In particular, the
expressions (\ref{ndpsc}) and (\ref{pdpsc}) for the steady state
density and survival probabilities will also apply. However, unlike
DP, we will see that the $\beta_{1,{\rm seed}}$ and $\beta_{1,{\rm
dens}}$ exponents are no longer equal. 

\subsubsection{Mean field theory for surface BARW}
\label{mfbarw}
\begin{figure}
\begin{center}
\leavevmode
\vbox{
\epsfxsize=45mm 
\epsffile{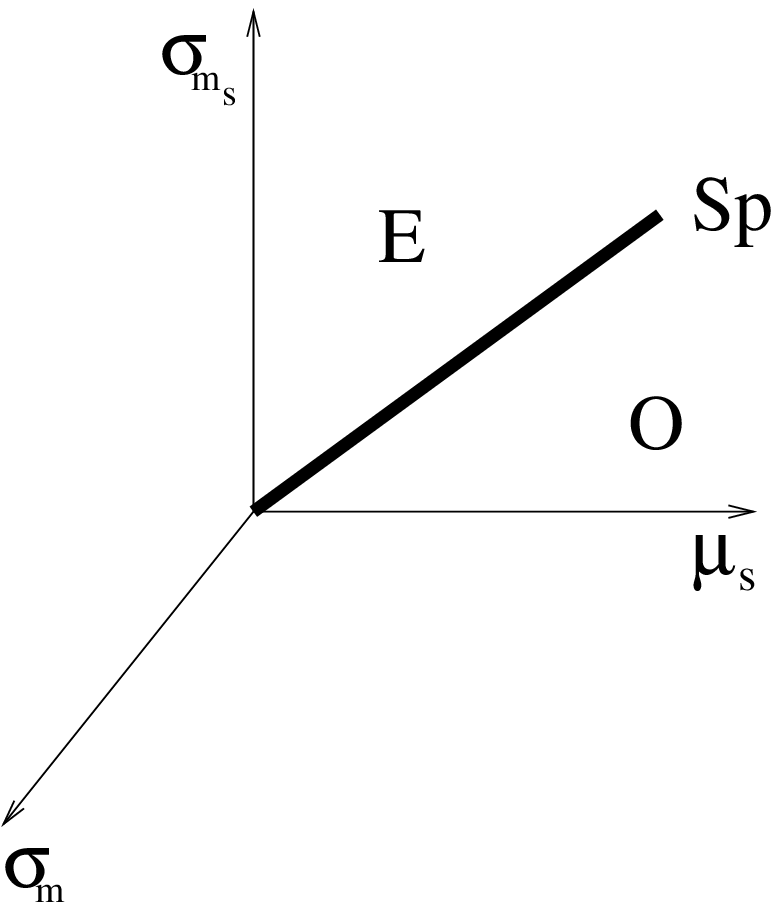}}
\end{center}
\vspace*{6mm}
\fcaption{Schematic mean field boundary phase diagram for BARW. See text
for an explanation of the labeling.} 
\label{psurf}
\end{figure}

\noindent
The surface phase diagram for the mean field theory of BARW (valid for
$d>d_c=2$) is shown in Figure~\ref{psurf}. Here $\sigma_m$,
$\sigma_{m_s}$ are the rates for the branching processes $A\to (m+1)A$
in the bulk and at the
surface, respectively, and $\mu_s$ is the rate for the surface
spontaneous annihilation reaction $A\to\emptyset$. Otherwise, the
labeling is the same as that for the DP phase diagram (see
Figure~\ref{dpps}). We summarize the main details of the phase diagram
below; more details can be found in Ref.\ \citeo{ourpre}.

The first feature to note is that the bulk is either active
($\sigma_m>0$) or critical ($\sigma_m=0$), but never inactive. Hence, 
unlike DP, there is no possibility of finding a surface transition,
where the surface is critical with the bulk inactive. For the case
where $\sigma_m=\mu_s=0$, we expect that for any finite value of the
surface branching, the surface will become active. This corresponds to
the extraordinary transition with an active surface and critical
bulk. On the other hand for $\sigma_m=\sigma_{m_s}=0$ and $\mu_s>0$, the
density at an (isolated) surface would decay away exponentially
quickly due to the $A\to\emptyset$ reaction, and hence we have the
ordinary transition. Consequently with $\sigma_m=0$,
but both $\mu_s$ and $\sigma_{m_s}$ non-zero, there should be a line of
special transitions dividing the extraordinary and ordinary
regions. This explains the general features of the phase diagram in
Figure~\ref{psurf}. 

At a more quantitative level, the mean field equation for BARW is very
similar to that for DP:
\begin{equation}
	\label{barwmfteq}
	\partial_t n = D\nabla^2 n -\Delta n -\lambda n^2 ,
\end{equation}
with the boundary condition 
\begin{equation}
	D \partial_{x_{\perp}}n|_{x_{\perp}=0}= \Delta_s
	n|_{x_{\perp}=0}. \label{eq:BCbarwgeneral}
\end{equation}
However the values of the $\Delta$, $\Delta_s$ parameters are now
different: $\Delta=-m\sigma_m$ and $\Delta_s=-m\sigma_{m_s}+\mu_s$.
For simplicity we will again focus only on the ordinary transition
(details of the other transitions can be found in Ref.\ \citeo{ourpre}).  
Clearly we expect the same mean field exponent $\beta_{1,{\rm dens}}$
as in DP.\cite{mistake}
However, the mean field behavior of the $\beta_{1,{\rm seed}}$
exponent is very different from the corresponding behavior in
DP. Consider placing two particles next to the surface at
$t=0$. From the recurrence properties of
random walks we see that, regardless of the reaction rates on the
surface or in the bulk, there is a finite chance that the two
particles will never meet again. Hence the survival probability is
{\it nonzero} and thus $\beta_{1,{\rm seed}}^{\rm O}=0$ in mean
field theory. 

\break

\subsection{Beyond mean field theory}
\label{d=1}
\vglue 0pt
\subsubsection{Surface DP}
\vglue 0pt
\noindent
We expect that the phase diagram shown in Figure~\ref{dpps} is
generally valid for surface DP close to the upper critical dimension
$d_c=4$. However, as first pointed out in Ref.\ \citeo{ourpre},
in $d=1$, where the surface is just a zero
dimensional point, the phase diagram may look rather different. In
that case, for an inactive bulk, net particle production
is only possible at one point. Furthermore, since
particles will be constantly lost into the bulk,
where they will decay away exponentially quickly, it may
not be possible to form an active surface state. Of course this
conclusion assumes the
absence of a surface reaction $\emptyset\to A$, which is the analog
of a surface magnetic field in the Ising model. If this reaction is
included then a normal transition obviously becomes possible.

\subsubsection{Boundary condition classification for $d=1$ DP cellular
automata}
\label{dpira}
\noindent
Following the analysis in the previous section, we can now attempt to assign 
universality classes for the IBC, RBC, ABC cellular automata boundary
conditions introduced earlier in Section~\ref{surfmodels}. The ABC
condition obviously behaves as
if there existed a surface reaction equivalent to $\emptyset\to A$, and
thus it belongs to the normal transition universality class.
Numerically, the IBC, RBC, belong to the same universality
class,\cite{ourprl} which we identify as the ordinary transition.

\subsubsection{Surface BARW}
\begin{figure}
\begin{center}
\leavevmode
\vbox{
\epsfxsize=3in
\epsffile{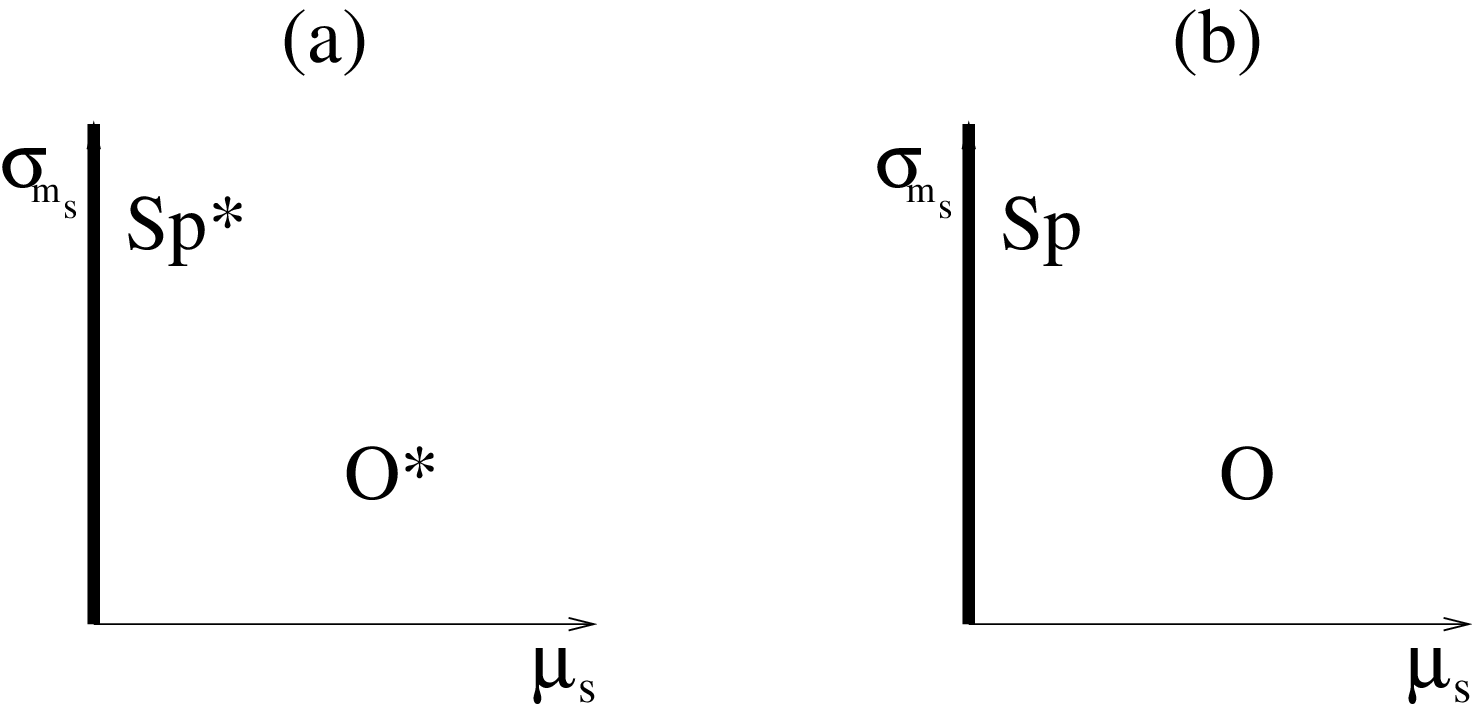}}
\end{center}
\vspace*{6mm}
\fcaption{Schematic surface phase diagrams for BARW in $d=1$ for (a)
$\sigma_m<\sigma_{m,{\rm critical}}$, and (b) $\sigma_m=\sigma_{m,{\rm
critical}}$. See text for an explanation of the labeling.} 
\label{psurf1d}
\end{figure}

\noindent
Next we turn our attention to the $d=1$ phase diagram for
surface BARW shown in Figure~\ref{psurf1d}. The phase diagram looks
quite different from its mean field analog due in part to the shift of
the bulk critical point away from zero branching rate, but also due to
the absence of any extraordinary transition. Again, in the
absence of the surface reactions of the form
$\emptyset\to A$, this is due to the fact that excess particle
production (with a finite reaction rate) at a zero dimensional surface
is simply not efficient enough to generate an active state, due to
leakage into the critical bulk.

We now divide the phase diagram into two sections, as in
Figure~\ref{psurf1d}. 

\

$\sigma_m<\sigma_{m,{\rm critical}}$: 

\

In this case, the bulk will be controlled by the annihilation process
$A+A\to\emptyset$, and the branching process will be everywhere
irrelevant. Consider first the case with $\mu_s=0$, where
the surface reaction $A\to\emptyset$ is not permitted. In that case,
since both the wall and bulk will be controlled by the critical
$A+A\to\emptyset$ reaction, we expect an analog of
the special transition. However, this will not be a special transition
in the usual sense, since in this region it will {\it not} be possible
to obtain an active state either on the surface or in the bulk by
small changes in the bulk and/or surface branching rates. To emphasize
this point, we label this regime as Sp* in Figure~\ref{psurf1d}a.
This simpler and analytically tractable case has already been
extensively analyzed in Ref.\ \citeo{magnus} (see also
Section~\ref{ibcfth}). Next we consider the case $\mu_s>0$, where 
the reaction $A\to\emptyset$ is permitted on the wall. Since the
branching is irrelevant in this region of parameter space,
the reaction $A\to\emptyset$ would give rise to an exponentially
decaying density at an isolated boundary, and hence an inactive state.
Consequently we expect an analog of
the ordinary transition. However, for the same reasons as described
above, we will label this regime as O* in Figure~\ref{psurf1d}a. 

\

$\sigma_m=\sigma_{m,{\rm critical}}$:

\

Since we are now at the bulk critical point, which borders the bulk
active phase, we expect rather different behavior to that described
above. In this case it {\it will} be possible to obtain an active
bulk/surface state by small changes in the reaction parameters. For
the case where $\mu_s=0$, with both the bulk and surface
critical, we expect a standard special transition (labeled as Sp in
Figure~\ref{psurf1d}b). On the other hand, if $\mu_s>0$, then the
parity symmetry of BARW is broken at the boundary and we expect a
critical bulk and (if isolated) an inactive surface, i.e. a standard
ordinary transition (labeled as O in Figure~\ref{psurf1d}b). 

\subsubsection{Boundary condition classification for $d=1$ BARW
cellular automata}
\label{barwira}
\noindent
We can now discuss the relations between the boundary conditions for
the $d=1$ DP2 cellular automata
introduced in Section~\ref{surfmodels} and our above classification of the
boundary universality classes for BARW in $d=1$. The
key feature is whether the symmetry in DP2 between the two absorbing
states in the bulk is preserved at the surface. This symmetry is the
analog of the parity symmetry conservation in the reaction-diffusion
BARW model, which allows the DP universality class to be escaped.
As we discussed above, breaking the parity symmetry on the boundary
leads to the ordinary transition. The analog of that situation is
provided 
by the RBC model which breaks the symmetry between the absorbing
states on the boundary and hence also belongs to the ordinary
transition. On the other hand, the IBC model respects the symmetry at
the boundary, and thus belongs to the
special transition. Furthermore, the ABC model once again
clearly belongs to the normal transition universality class. Hence, as
claimed earlier in Section~\ref{surfmodels}, we see that by
using the IBC, RBC, ABC classification all the previously discussed
boundary BARW transitions in $d=1$ can be accessed. 

\section{Scaling Theory}
\label{scalingtheory}
\noindent
In this section we review the scaling theory for the survival
probabilities and correlation functions for boundary DP and
BARW. For simplicity we again
restrict ourselves to the ordinary transition; additional details
can be found in Refs.\ \citeo{dp-wall-edge,ourprl,ourpre}.

\subsection{Surface DP}
\label{scthdp}
\noindent
As we mentioned in Section\ \ref{sdpgendisc}, the most important point
to emphasize is that there is only one
boundary $\beta_1^{\rm O}$ exponent for DP:
$\beta_1^{\rm O}=\beta_{1,{\rm seed}}^{\rm O}=\beta_{1,{\rm dens}}^{\rm O}$.
The survival probability (the probability that the cluster is still alive 
at time $t$) has the form
\begin{equation}
	P_1(t,\Delta) = \Delta^{\beta_1^{\rm O}} \, \psi_1 \left({{t}/
	{\xi_{\parallel}}}\right),            \label{eq:P_1(t,Delta)}
\end{equation}
where the scaling function $\psi_1$ is constant for 
$t \gg \xi_\parallel$.\cite{grassberger-torre}
The mean lifetime of finite clusters 
\begin{equation}
	\left< t \right> \sim |\Delta|^{-\tau_1^{\rm O}}
\label{taudef1}
\end{equation}
then follows from the lifetime distribution $-d P_1/dt$,\cite{essam-etal:1996}
yielding the exponent
\begin{equation}
	\label{tau1_relation}
	\tau_1^{\rm O} = \nu_\parallel - \beta_1^{\rm O}.
\end{equation}
However, for $\nu_\parallel < \beta_1^{\rm O}$, the leading contribution to 
$\langle t \rangle$ will be a constant, such that the above scaling relation 
breaks down and is replaced by $\tau_1 = 0$.

For the coarse-grained bulk density $n_1$ at the point (${\bf x}$, $t$)
of a cluster grown from a single seed located next to the wall at ${\bf x}=
{\bf 0}$, $t=0$, one can make the scaling ansatz
\begin{equation}
	\label{DPansatz_rho_wall}
	n_{1}(x,t,\Delta) = \Delta^{\beta_1^{\rm O} + \beta}
 	f_1  \left(x/{\xi_\perp}, \, {t}/{\xi_\parallel}\right).
\end{equation}
This ansatz may be properly justified using the field theory analysis
reviewed in Section \ref{Dpft}, 
however a more intuitive justification of the
prefactors may be given as follows.
The first factor of $\Delta^{\beta_1^{\rm O}}$ comes from the probability 
that an infinite cluster can be grown from the seed. 
The second factor of $\Delta^\beta$ arises from the bulk scaling of
activity in the active state, i.e., 
the (conditional) probability that the point (${\bf x}$, $t$) belongs to the
infinite cluster grown from the seed
(see also Ref.\ \citeo{grassberger-3d}). 
In contrast, if the density is measured at the wall, 
then the appropriate ansatz reads
\begin{equation}
	\label{DPansatz_rho_wall_wall}
	n_{11}(x,t,\Delta) = \Delta^{2\beta_1^{\rm O}}
  	f_{11} \left(x/{\xi_\perp}, \,
  	t/{\xi_\parallel}\right) ,
\end{equation}
as we pick up a factor $\Delta^{\beta_1^{\rm O}}$ rather than $\Delta^\beta$
for the probability that (${\bf x}$, $t$) at the wall belongs to the
infinite cluster grown from the seed.

By integrating the cluster density (\ref{DPansatz_rho_wall}) over space and
time, we arrive at the average size of finite clusters grown from
seeds on the wall,
\begin{equation}
	\label{DPsize_wall}
	\langle s \rangle \sim |\Delta|^{-\gamma_1^{\rm O}} ,
\end{equation}
such that
\begin{equation}
	\nu_{\parallel}+d\nu_{\perp}=\beta_1^{\rm O}
	+\beta+\gamma_{1}^{\rm O} .
	\label{DPsurf_hyperscaling}
\end{equation}
Hence, the surface exponent $\gamma_1$ is related to the previously
defined exponents via a scaling law that naturally generalizes the usual
$d+1$ dimensional bulk hyperscaling relation
\begin{equation}
	\nu_{\parallel} +d\nu_{\perp} = 2\beta+\gamma .
	\label{DPbulk_hyperscaling}
\end{equation}
The results of numerical simulations with a 
wall (see Section~\ref{sec:numres}) are in very good agreement 
with the hyperscaling relation (\ref{DPsurf_hyperscaling}).

Besides integrating the density (\ref{DPansatz_rho_wall}),
we can also integrate
the density on the wall (\ref{DPansatz_rho_wall_wall}) over the $d-1$
dimensional wall and time. This integration yields the average
(finite) cluster size on the wall,
\begin{equation}
	\label{DPsize_wall_wall}
	\langle s_{1,1} \rangle \sim |\Delta|^{-\gamma_{1,1}^{\rm O}} ,
\end{equation}
where
\begin{equation}
	\nu_{\parallel}+(d-1)\nu_{\perp}=2\beta_1^{\rm O}+
	\gamma_{1,1}^{\rm O} .
	\label{DPsurf_hyperscaling_wall_wall}
\end{equation}
However, in higher dimensions ($d \approx 2$ being a marginal case)
this relation is not fulfilled as it would predict a negative
$\gamma_{1,1}^{\rm O}$. For this case, $\gamma_{1,1}^{\rm O} = 0$,
reflecting a constant 
contribution to Eq.~(\ref{DPsize_wall_wall}), cf. the comment 
after Eq.~(\ref{tau1_relation}). 


\break

\subsection{Edge DP}\label{DPedge}
\noindent
We next turn to briefly review the case of DP clusters 
started on an edge. It has been known for some time that
the presence of an edge introduces new exponents, independent of those 
associated with the bulk or with a surface (see Ref.\ \citeo{cardy-edge}
for a discussion in the context of equilibrium critical phenomena, or 
Ref.\ \citeo{grassberger-3d} in the context of percolation). 

Consider a system, where we allow the wall to have an edge with an angle 
$\alpha$ at $x_{\parallel}^{(1)}=x_\perp=0$. Hence, the edge is simply
the $d-2$ dimensional intersection of two $d-1$ dimensional
walls. By placing the seed next to this edge, the boundary exponent 
$\beta_1^{\rm O}$ is replaced by the edge exponent $\beta_2^{\rm O}(\alpha)$ 
(where of course $\beta_1^{\rm O} = \beta_2^{\rm O}(\pi)$). 
Following the same arguments as before, 
we have the new scaling ansatz for the cluster density
\begin{equation}
\label{ansatz_rho_edge}
n_{2}(r,t,\Delta) = 
 	\Delta^{\beta_2^{\rm O} + \beta} 
 	f_2 \left( {r}/{\xi_\perp}, \, {t}/{\xi_\parallel}\right) ,
\end{equation}
where $r$ is the radial coordinate in a system of spherical polar
coordinates centered on $x_{\parallel}^{(1)}=x_\perp=0$.
This ansatz applies for directions away from the edge
and the walls. By replacing $\beta$ with $\beta_1^{\rm O}$ or
$\beta_2^{\rm O}$, we get the
corresponding results for the density along the wall or the edge,
respectively. Moreover, in analogy with Eqs.~(\ref{DPsize_wall}) and 
(\ref{DPsurf_hyperscaling}) for seeds on a wall, we obtain the average 
(finite) size $\langle s \rangle \sim |\Delta|^{-\gamma_2^{\rm O}}$ of
clusters grown from a seed next to an edge, by integrating
Eq.~(\ref{ansatz_rho_edge}) over space and time.
This yields the hyperscaling relation
\begin{equation}
	\nu_{\parallel}+d\nu_{\perp}=\beta_2^{\rm
	O}+\beta+\gamma_{2}^{\rm O} . 
	\label{edge_hyperscaling}
\end{equation}
Quoting the results from Ref.\ \citeo{dp-wall-edge} for a
mean field calculation of the edge exponents
\begin{equation}
	\gamma_2^{\rm O}=1-\pi/2\alpha ,
        \quad \beta_2^{\rm O}=1+\pi/ 2\alpha ,
	\label{eq:gamma2}
\end{equation}
we see that Eq.~(\ref{edge_hyperscaling}) is satisfied at
the upper critical dimension $d_c=4$. 
Numerical estimates for the exponent $\beta_2^{\rm O}$ from
Ref.\ \citeo{dp-wall-edge}, together with the mean field
values, are listed in Table~\ref{table-beta}
(see Section~\ref{sec:numres} for further details on simulation
methods). 

\begin{table}[htb]
\tcaption{Numerical estimates for the $\beta_2^{\rm O}$ exponents for edge DP 
        together with the mean field values from Eq.~(\protect\ref{eq:gamma2}).
         Recall that $\beta_2^{\rm O}(\pi)=\beta_1^{\rm O}$.
	   The bulk and $d=1$ wall estimates are listed 
	   for reference.
}
\vspace*{0.5cm}
\centerline{\footnotesize\smalllineskip
\begin{tabular}{l|c|c|c|c|c} 
Angle ($\alpha$)  &  \makebox[16mm]{$\pi/2$}   & \makebox[16mm]{$3\pi/4$}  
                  &  \makebox[16mm]{$\pi$}     & \makebox[16mm]{$5\pi/4$}  
                  &  \makebox[16mm]{bulk} \\ \hline
${\beta_{1}^{\rm O}}_{~}^{~}~(d=1)$  & & & $0.733 71(2)$  & & $0.276 486(8)$ \\
${\beta_{2}^{\rm O}}_{~}^{~}~(d=2)$  & $1.6(1)$    & $1.23(7)$   
                     & $1.07(5)$ & $0.98(5)$ & $0.583(4)$\\ 
${\beta_{2}^{\rm O}}_{~}^{~}~({\rm MF})$ 
                      &    2    &   5/3  &   3/2   &   7/5   &   1  \\
\end{tabular}
}
\vspace*{0.5cm}
\label{table-beta}
\end{table}

\subsection{Surface BARW}\label{corrbarw}
\label{scthbarw}
\vglue 0pt
\noindent
In this section we briefly review
the analogous scaling theory for surface BARW.  Again, for brevity, we will
only consider the ordinary transition. When writing down this scaling 
theory we must now bear in mind the
important distinction between the $\beta_{1,{\rm dens}}$ and
$\beta_{1,{\rm seed}}$ exponents. 
For a seed placed on the wall at ${\bf x}={\bf 0}$, $t=0$,
the scaling form for the survival probability has the form 
\begin{equation}
	\label{barwsurvwall}
	P_1(t,\Delta)=|\Delta|^{\beta^{\rm O}_{1,{\rm seed}}}
	\Phi_1(t/\xi_{\parallel}). 
\end{equation}
It is then straightforward to compute the average lifetime of finite
clusters, $\langle t \rangle \sim |\Delta|^{-\tau_1}$, where
$\tau_1^{\rm O} = \nu_\parallel - \beta_{1, \rm seed}^{\rm O}$,
just as in the case of DP.  

For the coarse-grained bulk particle density 
$n_1$ at the point (${\bf x}$, $t$) of a cluster grown
from a single seed located next to the wall at ${\bf x}={\bf 0}$, $t=0$,
one can make the ansatz
\begin{equation}
        \label{ansatz_rho_wall}
        n_{1}(x,t,\Delta) = |\Delta|^{\beta^{\rm O}_{1, \rm seed} +
        \beta_{\rm dens}}
        g_1  \left(x/{\xi_\perp}, \, {t}/{\xi_\parallel}\right) .
\end{equation}
As was the case for DP, the $\Delta$-prefactor in
Eq.~(\ref{ansatz_rho_wall}) comes from  
Eq.~(\ref{barwsurvwall}) for the probability that an infinite cluster can be
grown from the seed, and from Eq.~(\ref{n(Delta)}) for the (conditional)
probability that the point (${\bf x}$, $t$) belongs to this cluster.
If, instead,  the density is measured at the wall, we have
\begin{equation}
        \label{ansatz_rho_wall_wall}
        n_{11}(x,t,\Delta) =
        |\Delta|^{\beta^{\rm O}_{1, \rm seed} + \beta^{\rm O}_{1, \rm dens}}
        g_{11} \left(x/{\xi_\perp}, \,
        t/{\xi_\parallel}\right) ,
\end{equation}
as we pick up a factor $|\Delta|^{\beta^{\rm O}_{1, \rm dens}}$ rather
than $|\Delta|^{\beta_{\rm dens}}$ from the probability that (${\bf x}$,
$t$) at the wall belongs to the cluster. 
The average size of finite clusters 
\begin{equation}
        \label{size_wall}
        \langle s_1 \rangle \sim |\Delta|^{-\gamma^{\rm O}_1} ,
\end{equation}
follows from integrating the cluster density 
(\ref{ansatz_rho_wall}) over space and time,
where the exponent $\gamma^{\rm O}_1$ is
related to the previously defined exponents via 
\begin{equation}
        \nu_{\parallel}+d\nu_{\perp}=
        \beta^{\rm O}_{1, \rm seed}+\beta_{\rm dens}+\gamma^{\rm O}_{1} .
        \label{surf_hyperscaling}
\end{equation}
Analogously, by integrating the cluster wall density 
(\ref{ansatz_rho_wall_wall}) over the ($d-1$)-dimensional wall and 
time, we obtain the average size of finite clusters on the wall
\begin{equation}
        \label{size_wall_wall}
        \langle s_{1,1} \rangle \sim |\Delta|^{-\gamma^{\rm O}_{1,1}} ,
\end{equation}
where
\begin{equation}
        \nu_{\parallel}+(d-1)\nu_{\perp}=
          \beta^{\rm O}_{1, \rm seed}+\beta^{\rm O}_{1, \rm dens}+
          \gamma^{\rm O}_{1,1}.
        \label{surf_hyperscaling_wall_wall}
\end{equation}
Note that if the $\gamma$ exponents obtained from 
Eqs.~(\ref{surf_hyperscaling}) and (\ref{surf_hyperscaling_wall_wall})
are negative, then they should be replaced by zero in Eqs.~(\ref{size_wall}) 
and (\ref{size_wall_wall}).

\section{Field Theory}\label{sec:ft}
\vglue 0pt
\noindent
In the following sections we briefly review the available field theoretic
results for bulk and boundary\cite{janssen-etal,ourpre} DP and BARW.
These techniques allow for the systematic inclusion of fluctuation effects,
important below the upper critical dimension $d_c$.

\break

\subsection{DP field theory}\label{Dpft}
\noindent
The field theory describing bulk DP is
very well known.\cite{cardy-sugar,sundermeyer}  The action is given by
\begin{equation}
	S_{\rm bulk}=\int d^dx\int dt ~ \left(\, \bar\phi\, [\, \partial_t
	-D\nabla^2 +\Delta \, ]\,\phi
	+{1\over 2}u\, [\, \bar\phi\phi^2-\bar\phi^2\phi\, ]\,
	\right),
	\label{actionbulk}
\end{equation}
where $\phi({\bf x},t)$ is the ``density'' field,
and where $\bar\phi({\bf x},t)$ is the
response field. Simple power counting reveals
that the upper critical dimension is $d_c=4$ below which fluctuation
effects become important. Renormalization
of the theory is standard,
involving field, mass ($\Delta$), diffusion constant 
($D$) and coupling constant ($u$) renormalizations.
The critical exponents can then be computed perturbatively
using an $\epsilon=4-d$ expansion, giving\cite{janssen1981}
\begin{equation}
	\beta_{\rm seed}=
	\beta_{\rm dens}=1-{\epsilon\over 6}+O(\epsilon^2), \quad
	\nu_{\parallel}= 1+{\epsilon\over 12}+O(\epsilon^2), \quad
	\nu_{\perp}= {1\over 2}+{\epsilon\over 16}+O(\epsilon^2).
\end{equation} 
Note that, in this case,
the exponents $\beta_{\rm seed}$ and $\beta_{\rm dens}$ can
be shown to be equal.
Technically, this follows from the fact that the density 
and response fields renormalize identically (see also
Ref.\ \citeo{mgt}).  

In the action~(\ref{actionbulk}) one can integrate out the response
field $\bar\phi({\bf x},t)$ and arrive at a Langevin equation for the
local density $\phi({\bf x},t)$:
\begin{equation}
        \frac{\partial \phi({\bf x},t)}{\partial t} =
        D\nabla^2\phi({\bf x},t) - \Delta \phi({\bf x},t)
        -{1\over 2}u\phi({\bf x},t)^2 +\eta({\bf x},t)  , 
                                \label{langevin} 
\end{equation}
with
\begin{equation}
        \langle\eta({\bf x},t)\rangle=0, \qquad 
        \langle\eta({\bf x},t)\eta({\bf x}',t')\rangle
        = u\phi({\bf x},t)\delta^d({\bf x}-{\bf x}')\delta(t-t'),
                                        \label{langevin-noise}
\end{equation}
where $\eta({\bf x},t)$ is a Gaussian noise term. The multiplicative
factor $\phi({\bf x},t)$ in the noise correlator reflects the fact that
$\phi=0$ is the absorbing state. 

The use of field theories to study boundary nonequilibrium phase transitions
was initiated by Janssen et al.\cite{janssen-etal}  They showed that the 
appropriate action for DP with a wall at $x_{\perp}=0$ is given by
$S=S_{\rm bulk}+S_{\rm surface},$ where
\begin{equation}
	S_{\rm surface}= \int d^{d-1}x\int dt ~ \Delta_s\,
	\bar\phi_s\, \phi_s, 
	\label{actionsurface}
\end{equation}
with the definitions
	$\phi_s=\phi ({\bf x_{\parallel}},x_{\perp}=0,t)$ and 
	$\bar \phi_s =\bar\phi({\bf x_{\parallel}},x_{\perp}=0,t)$.
The surface term $S_{\rm surface}$
corresponds to the most relevant interaction consistent with the 
symmetries of the problem and which also respects the absorbing state
condition. Simple power counting indicates that
$[\Delta_s]\sim\kappa$, where $\kappa$ denotes an inverse length
scale. The presence of the wall implies the boundary condition at
$x_{\perp}=0$ of   
\begin{equation}
	\left. D\partial_{x_{\perp}} \phi\right|_s= \Delta_s\phi_s . 
\end{equation}
Using this boundary condition, we 
see that a boundary term of the form $\bar\phi_s\partial_{x_{\perp}}\phi_s$, 
although marginal from power counting arguments, is actually redundant.

Since $\Delta_s\sim\kappa$, its renormalized value can only flow to
one of three possible fixed points: $0$ or $\pm\infty$. These
possibilities correspond to the various types of boundary phase transitions. 
The flow $\Delta_s\to-\infty~(\infty)$ corresponds to the extraordinary
(ordinary) 
transition, whereas the unstable fixed point at $\Delta_s=0$ corresponds
to the multicritical special transition. In the following we will concentrate
solely on 
the ordinary transition, since that is the only case to have been studied 
numerically (see Ref.\ \citeo{janssen-etal} for a field theoretic
analysis of the special transition).

The only extra renormalization required by the presence of the wall is a
surface field renormalization.\cite{janssen-etal}  The presence of this extra
renormalization at the surface leads naturally to the existence of just one
independent surface exponent $\beta_1^{\rm O}$. Once again, the exponents 
$\beta_{1,{\rm seed}}^{\rm O}$ and $\beta_{1,{\rm dens}}^{\rm O}$ can
be shown to be equal\cite{dp-wall-edge} $\beta_{1,{\rm seed}}^{\rm
O}=\beta_{1,{\rm dens}}^{\rm O}= 
\beta_1^{\rm O}$, similar to the result found in the bulk. This
follows from the fact that the {\em surface} density and response fields
renormalize identically. Furthermore it can be shown that 
the correlation length exponents are everywhere unchanged by the wall 
(see Refs.\ \citeo{diehl,janssen-etal}).
The $\beta_1^{\rm O}$ exponent can again be computed using an $\epsilon=4-d$ 
expansion yielding\cite{janssen-etal}
\begin{equation}
	\label{beta_1}
	\beta_{1}^{\rm O}=\beta_{1,{\rm seed}}^{\rm O}=
        \beta_{1,{\rm dens}}^{\rm O}=
	{3\over 2}-{7\epsilon\over 48}+O(\epsilon^2). \\
\end{equation} 
{}From the field theory of Ref.\ \citeo{janssen-etal},
it is not hard to verify 
that Eq.~(\ref{DPsurf_hyperscaling}) is the appropriate generalization
of Eq.~(\ref{DPbulk_hyperscaling}), relating $\beta_1^{\rm O}$ to
\begin{equation}
	\gamma_{1}^{\rm O}={1\over 2}+{7\epsilon\over 48}+O(\epsilon^2).
\end{equation}

\subsection{BARW field theory}\label{BARWft}
\vglue 0pt
\noindent
We now briefly review the bulk field theory for the BARW
reaction-diffusion system. In 
order to properly include fluctuation effects for BARW, one must be 
careful to include processes generated by a combination of branching
and annihilation. In other words in addition to the process $A\to
(m+1)A$, the reactions $A\to (m-1)A$, \ldots, $A\to 3A$ need to be
included. These considerations lead to the full action\cite{cardy-tauber}
\begin{eqnarray}
	& & S_{\rm bulk}[\psi,\hat\psi;\tau]=
	\int d^dx \left[ \int_0^{\tau} dt \left( \hat\psi({\bf x},t)
	[\partial_t- D{\bf\nabla}^2]\psi({\bf x},t) \right.\right.
								\nonumber \\
	& & \left.\left. \,\;\;\;\qquad\qquad\qquad
	+ \sum_{l=1}^{m/2}\sigma_{2l}[1-\hat\psi({\bf x},t)^{2l}]
	  \hat\psi({\bf x},t)\psi({\bf x},t) \right.\right.
							\label{barwbulk} \\
	& & \left.\left. \quad\qquad\qquad\qquad
        - \lambda[1-\hat\psi({\bf x},t)^2]\psi({\bf x},t)^2 \right)
	  -\psi({\bf x},\tau)-n_0\hat\psi({\bf x},0) \right] , \nonumber 
\end{eqnarray}
written in terms of the response field
$\hat\psi({\bf x},t)$ and the ``density'' field $\psi({\bf x},t)$.
Here $D$ is the diffusion constant, $\lambda$ the annihilation rate, and
$\sigma_{2l}$ the branching rate for the process $A\to (2l+1)A$. Note that 
the final two terms in Eq.~(\ref{barwbulk}) represent, respectively,  
a contribution due to the projection state (see Ref.\ \citeo{peliti}),
and the  
initial condition (an uncorrelated Poisson distribution with mean $n_0$). 
Notice also that (for {\it even $m$}) the action (\ref{barwbulk})
is invariant under the ``parity'' transformation
\begin{equation}
\label{parity}
\hat\psi({\bf x},t)\to -\hat\psi({\bf x},t), \qquad \psi({\bf x},t)\to
-\psi({\bf x},t) .
\end{equation}
This symmetry corresponds physically to particle conservation modulo
$2$ and it enables the system 
to escape the DP universality class.

Simple power counting on the action in Eq.~(\ref{barwbulk}) reveals
that the upper critical dimension is $d_c=2$. Close to $d_c$,
the renormalization of the above action is
quite straightforward (here we again quote the results from 
Ref.\ \citeo{cardy-tauber}). At the annihilation fixed point the RG
eigenvalue of the branching parameter can easily be computed. To one
loop order one finds
$y_{\mu_m}=2-m(m+1)\epsilon/2+O(\epsilon^2)$, where $\epsilon=2-d$.
Hence we see that the {\it lowest} branching process is
actually the most relevant. Therefore, close to $2$ dimensions, where 
the branching remains relevant, we expect to find an {\it active}
state for all nonzero values of the branching (as was the case for
the BARW mean field theory reviewed earlier).

However, inspection of the above most relevant RG eigenvalue
$y_{\sigma_2}$ shows
that it eventually becomes negative at $d=d_c'\approx 4/3$. In that
case we expect a major change in the behavior of
the system, since the branching process will no longer be relevant at
the annihilation fixed point. The critical transition point is then
shifted with the active state only being present for values of the
branching greater than some positive critical value (as indicated in
Figure~\ref{pbulk}b).   
Consequently, we see that there is a second critical
dimension $d_c'\approx 4/3$ whose presence immediately rules out any
possibility of accessing the non-trivial behavior expected in $d=1$
via perturbative epsilon expansions down from $d=2$. Instead cruder
techniques (such as the loop expansion in fixed dimension) must be
employed.\cite{cardy-tauber} 

We now review the effects of introducing a surface into the BARW 
universality class\cite{ourpre} and further allow for the possibility
of a symmetry-breaking $A\to\emptyset$ reaction to take place
(with rate $\mu_s$), but only on the surface. 

\subsection{$\mu_s=0$ BARW field theory}
\label{ibcfth}
\vglue 0pt
\noindent
In this case only the branching process is relevant on
the surface. However, as in the bulk, one must still be careful to include 
the surface branching processes generated by a combination of branching and 
annihilation. This leads to a full surface action of the form
\begin{equation}
\label{IBCactionfull}
S_{\rm surface}= \int d^{d-1}x_{\parallel} \int_0^{\tau} dt ~ \left(
\sum_{l=1}^{m/2}\sigma_{2l_{s}}\left(1-\hat\psi_s^{2l}\right)
\hat\psi_s\psi_s \right) ,
\end{equation}
where $\hat\psi_s=\hat\psi({\bf x_{\parallel}},x_{\perp}=0,t)$ and
$\psi_s=\psi({\bf x_{\parallel}},x_{\perp}=0,t)$. 
Notice that the parity symmetry (\ref{parity}) is preserved
for the $\mu_s=0$ model at the wall, as well as in the bulk. This boundary
action leads to boundary conditions very similar to those discussed earlier
for DP, of the form $D\partial_{x_{\perp}}\psi|_{x_{\perp}}=\Delta_s\psi_s$,
where 
$\Delta_s$ is the surface mass, equal to $-m\sigma_{m_s}$ in mean field theory.

Power counting on the above action reveals that the
surface branching rates $\sigma_{2l_s}$ all have naive dimension
$[\sigma_{2l_s}]\sim \kappa^1$, where $\kappa$ denotes an inverse
length scale. However, close to, but below $d=2$, this scaling dimension
will be renormalized downwards (this can be seen physically as a
result of processes like $A\to 3A\to A$ rendering the
branching process less efficient). As a result of this
renormalization, we expect the lowest generated process
(i.e.\ with $l=1$ in Eq.~(\ref{IBCactionfull})) will become the {\it
most} relevant (as it was in the bulk). Nevertheless, despite this
downward renormalization, close enough to $d=2$,  
the scaling dimension of the most relevant coupling $\sigma_{2_s}$
will remain positive, and thus it will flow to $\infty$ for all nonzero 
starting values. 
This state of affairs corresponds to the extraordinary transition,
where the surface is {\it active} while the bulk is critical.
On the other hand, at bulk criticality and with $\sigma_{2_s}=0$, we
have a multicritical special transition point. Hence, for
$\mu_s=0$ and close to $d=2$ the basic structure of the phase diagram
is unchanged from mean field theory, although the values of the 
exponents will of course be altered by the fluctuations.
In this case, after writing down and solving the
appropriate RG equations (exactly along the lines of
Refs.\ \citeo{cardy-tauber,janssen-etal}), one can derive scaling results
for the density  similar to those quoted in Section~\ref{corrbarw}. 

The situation in $d=1$ is rather different, partly due to the
shift of the bulk critical point away from $\sigma_m=0$.
This ensures that the boundary and bulk transitions in $d=1$ are
inaccessible to controlled perturbative expansions.
Nevertheless we still expect the scaling
dimension of all the $\sigma_{2l_s}$ to be negative in $d=1$,
following the downwards trend in the renormalization mentioned
above. In that case surface branching is then {\it irrelevant} in
$d=1$ leading to the Sp* and Sp 
special transitions discussed earlier. Hence, if the above scenario is
correct,  
we do not expect to see an extraordinary transition in $d=1$ for any finite 
value of the surface branching, since the surface branching will always be 
irrelevant. This conclusion was investigated numerically in
Ref.\ \citeo{ourpre}, where no evidence of an active
surface state for $\sigma_m\leq\sigma_{m,{\rm critical}}$ 
was found even for very high values of the surface branching
parameter in a fermionic lattice model in $d=1$. 

\subsection{$\mu_s\neq 0$ BARW field theory}
\label{rbcfth}
\vglue 0pt
\noindent
In this case the reaction $A\to\emptyset$ {\it is} now possible, but
only at sites on the wall. Including surface processes generated by 
a combination of branching and annihilation (i.e. the processes
$A\to mA$, $A\to (m-1)A$, $\ldots$, $A\to 2A$), the full surface action
becomes
\begin{equation}
	S_2= \int d^{d-1}x_{\parallel} \int_0^{\tau} dt ~ \left(
		\sum_{l=1}^m  ~ \sigma_{l_s} (1-\hat\psi^l_s) \hat\psi_s\psi_s
		+ \mu_s(\hat\psi_s-1) \psi_s\right) .
						\label{RBCgenaction}
\end{equation}
Note that the symmetry (\ref{parity}) is now broken by the surface
term proportional to $\mu_s$, which describes the $A\to\emptyset$
reaction. Once again we have a boundary condition of the form
$D\partial_{x_{\perp}}\psi|_{x_{\perp}}=\Delta_s\psi_s$, where now
$\Delta_s=-m\sigma_{m_s}+\mu_s$ in mean field theory.

The renormalization of the action (\ref{RBCgenaction}) is now
somewhat different from  the $\mu_s=0$ case. We again expect that
we need only keep the lowest generated branching term on the surface,
namely that with $l=1$ in Eq.~(\ref{RBCgenaction}). 
As before, we expect fluctuations to lower the
scaling dimension of this coupling from its mean field
value (although actually in $d=2$ this suppression will only be
logarithmic).  On the other hand, the efficacy of the $A\to\emptyset$
reaction is certainly {\it not} reduced by fluctuations. Hence, for $d\leq 2$,
we expect that $\Delta_s\sim \mu_s-\sigma_{1_s}$ will always run to the
fixed point at $\infty$, corresponding to the ordinary transition.
In particular we expect this picture to also hold in $d=1$, although
in that case the transition at $\sigma_m=\sigma_{m,{\rm critical}}$
will not be accessible to controlled perturbative expansions.
In $d=1$ we therefore
expect to find the O* ($\sigma_m<\sigma_{m,{\rm critical}})$ or O 
($\sigma_m=\sigma_{m,{\rm critical}})$ transitions mentioned earlier. 
Hence we find a state of affairs very different from mean field theory:
fluctuation effects now ensure that only the ordinary transition
is accessible for $d\leq 2$, when $\mu_s>0$. 

\section{Exact Results}
\label{exres}
\vglue 0pt
\noindent
In the last section we reviewed the use of field theoretic methods to
understand the effects of fluctuations in the boundary
DP and BARW models. Unfortunately,
for the case of BARW, the presence of a second critical dimension
$d_c'$ prevents a controlled, perturbative investigation of the system
in $d=1$. Hence, given the fundamental difficulties associated with
the BARW field theory, it seems fruitful to search for alternative
approaches. One such alternative is provided by the theory of quantum
spin Hamiltonians to which we now turn.

The methods we review, first presented in Ref.\ \citeo{ourpre}, are a 
straightforward
extension of the work in Refs.\ \citeo{schutz2} and \citeo{schutz1}.
The starting point is the following set of rules for BARW with $m=2$
in $d=1$: 
\begin{eqnarray}
& \emptyset A \leftrightarrow A \emptyset & ~~{\rm with~rate}~D/2
\nonumber \\
\label{fbarwrules}
& A A \to \emptyset\emptyset & ~~{\rm with~rate}~\lambda \\
& \emptyset A \emptyset \leftrightarrow A A A~~{\rm and}~~\emptyset A A
\leftrightarrow A A \emptyset & ~~{\rm with~rate}~\alpha/2 . \nonumber
\end{eqnarray}
Note that these rules are fermionic in character (no more than one
particle per site is permitted) in contrast to the bosonic rules
employed in the derivation of the earlier field theories.\cite{peliti}
The model described in Eq.~(\ref{fbarwrules}) can be transformed into a
spin picture by writing the
configuration of a semi-infinite system as a vector
$|s_1,s_2,s_3,\ldots\rangle$, where $s_i=1/2$ if the $i$-th site is
empty, and $s_i=-1/2$ if that site is occupied. Hence the system ket is
given by
\begin{equation}
|P(t)\rangle=\sum_{\{s_i\}}P(\{s_i\};t)|\{s_i\}\rangle ,
\end{equation}
and the equation governing the time evolution is
\begin{equation}
\partial_t|P(t)\rangle=-{\cal H}|P(t)\rangle ,
\end{equation}
where, using a representation in terms of Pauli matrices, and defining
$n_k=(1-\sigma_k^z)/2$, $v_k=1-n_k$, $s_k^{\pm}=(\sigma_k^x\pm
i\sigma_k^y)/2$, we have\cite{schutz2}
\begin{eqnarray}
 & & {\cal H}={1\over 2}\sum_{k=1}^{\infty}\left(D[n_kv_{k+1}+v_kn_{k+1}-
s_k^+s_{k+1}^- - s_k^-s_{k+1}^+] \right. \nonumber \\
 & & \left. ~~~~~~ + \, 2\lambda[n_kn_{k+1}-s_k^+s_{k+1}^+]
\right)+{\alpha\over 2}\sum_{k=2}^{\infty}(1-\sigma_{k-1}^x
\sigma_{k+1}^x)n_k \nonumber \\
 & & \quad =DH^{\rm SEP}+\lambda H^{\rm RSA}+\alpha H^{\rm
BARW} \label{spinh} \\
 & & \quad =D\sum_{k=1}^{\infty}h_k^{\rm SEP}+\lambda
\sum_{k=1}^{\infty}h_k^{\rm 
RSA}+\alpha\sum_{k=2}^{\infty}h_k^{\rm BARW} . \nonumber
\end{eqnarray}
Here we have used some of the notation of Ref.\ \citeo{schutz2}, where
SEP (symmetric exclusion process) refers to the diffusion piece, RSA
(random-sequential adsorption) to the annihilation piece and BARW
to the branching piece of the ``quantum Hamiltonian''. Notice that 
the boundary has been included in Eq.~(\ref{spinh}), since particles may
not hop to the left of site $1$, and the
annihilation/branching processes have also been restricted to sites
$1,2,3\ldots$. Hence, the above operator ${\cal H}$ governs the
evolution of a $d=1$ BARW system {\it without} an
$A\to\emptyset$ reaction at the 
boundary. Averages are calculated using the projection state
$\langle|=\sum_{\{s_i\}} \langle\{s_i\}|$, i.e. $\langle {\cal
F}\rangle = \langle | {\cal F}|P(t)\rangle$.
Following Ref.\ \citeo{schutz2}, we now define an operator ${\cal D}$
where 
\begin{equation}
{\cal D}=\gamma_{-1}\gamma_0\gamma_1\gamma_2\ldots ,
\end{equation}
with
\begin{eqnarray}
& & \gamma_{2k-1}={1\over 2}[(1+i)\sigma_k^z-(1-i)] , \nonumber \\
& & \gamma_{2k}={1\over 2}[(1+i)\sigma_k^x\sigma_{k+1}^x-(1-i)] .
\end{eqnarray}
Defining a new ``quantum Hamiltonian'' as $\tilde{\cal
H}=[{\cal D}^{-1}H{\cal D}]^T$, we find
\begin{eqnarray}
& & \tilde{\cal H}=(D-\lambda)\sum_{k=1}^{\infty}h_k^{\rm
BARW}+(\alpha+\lambda)\sum_{k=1}^{\infty}h_k^{\rm SEP}+\lambda
\sum_{k=1}^{\infty}h_k^{\rm RSA} \nonumber \\
& & \qquad\qquad\qquad+{\lambda\over
2}(n_1n_0-s_1^+s_0^++n_1v_0-s_1^+s_0^-) ,
\end{eqnarray}
where we have used the commutation rules described in detail in
Ref.\ \citeo{schutz2}. Hence, when $D=\lambda+\alpha$, we have the
following processes occurring:  
\begin{eqnarray}
& \emptyset_i A_{i+1} \leftrightarrow A_i \emptyset_{i+1} & ~{\rm
rate} ~(\lambda+\alpha)/2,\qquad\:i=1,2,3,\ldots, \nonumber \\
& A_i A_{i+1} \to \emptyset_i\emptyset_{i+1} & ~{\rm
rate}~\lambda,\;\;\;\;\;\;\qquad\qquad\:i=1,2,3,\ldots, \nonumber \\
& \emptyset_{i-1} A_i \emptyset_{i+1} \leftrightarrow A_{i-1} A_i
A_{i+1} 
& ~{\rm rate}~\alpha/2,\;\;\;\;\;\;\qquad\,~~\:\, i=1,2,3,\ldots, \nonumber\\
& \emptyset_{i-1}A_iA_{i+1}\leftrightarrow A_{i-1}A_i\emptyset_{i+1}
& ~ {\rm rate}~\alpha/2,\;\;\;\;\;\;\qquad\,~~\;\,i=1,2,3,\ldots, \nonumber \\
& A_0 A_1 \to \emptyset_0\emptyset_1 & ~{\rm rate}~\lambda/2,
\nonumber \\
& \emptyset_0 A_1 \to A_0 \emptyset_1 & ~{\rm rate}~\lambda/2 .
\label{edgeprocess}  
\end{eqnarray} 
Excepting the boundary terms, we see that the Hamiltonian has been
mapped back onto itself. Furthermore, at the edge, the particles may
only hop from site $1$ to site $0$
but never the other way round. This means that we can forget
about the $0$-th site in exchange for allowing the
processes $A_1 A_2\leftrightarrow A_1 \emptyset_2$ (with rate
$\alpha/2$), and $A_1\to\emptyset_1$
(with rate $\lambda/2$). Hence, we see that the new Hamiltonian
$\tilde{\cal H}$ corresponds to the case where $\mu_s\neq 0$, with the
DP processes $A\leftrightarrow A+A$ and $A\to\emptyset$ generated on
the boundary. 

If we choose the initial condition to be an uncorrelated state
with density $1/2$, denoted by $|1/2\rangle$, then the density at site
$k$, $\rho_k(t)$, is given by
\begin{equation}
\label{exeq}
\rho_k(t)=\langle|n_k\exp(-{\cal H}t)|1/2\rangle .
\end{equation}
Following exactly the procedure in Refs.\ \citeo{schutz2} and
\citeo{schutz1} (starting
with insertions of the identity operator ${\cal D}{\cal D}^{-1}$ into
the RHS of Eq.~(\ref{exeq})), one can straightforwardly show that
\begin{equation}
\label{equiv}
\rho_k(t)={1\over 2}[1-\langle 0|\exp(-\tilde{\cal H}t)|k-1,k\rangle] ,
\end{equation}
where $\langle 0|$ is the vacuum state (with no particles), and
$|k-1,k\rangle$ is the initial state with only two particles situated
at sites $k-1$ and $k$. In Eq.~(\ref{equiv})
the LHS is the density at the $k$-th site,
whereas the RHS is one-half times the probability that a cluster
initiated at $t=0$ by two particles
at sites $k-1$ and $k$ has not yet died out by time
$t$. According to our earlier analysis, for $\Delta<0$, the LHS should
scale as 
$|\Delta|^{\beta_{\rm dens}}$ (far from the wall) or
$|\Delta|^{\beta^{\rm Sp}_{1,{\rm dens}}}$ (close to the wall), and
the RHS as $|\Delta|^{\beta_{\rm seed}}$ (far from the wall) or
$|\Delta|^{\beta^{\rm O}_{1,{\rm seed}}}$ (close to the wall). Thus,
at the line $D=\lambda+\alpha$, we have proven the result
\begin{equation}
	\beta^{\rm O}_{1,{\rm seed}}=\beta^{\rm Sp}_{1,{\rm dens}}
				\label{duality1}
\end{equation}
(and, of
course, the bulk result $\beta_{\rm seed}=\beta_{\rm dens}$). We note
that the bulk result was first proven in Ref.\ \citeo{schutz2}, and a
very similar result for the $A+A\to\emptyset$ reaction was derived in
Ref.\ \citeo{schutz1} (connecting
the O* and Sp* regimes).
Using universality, it was postulated in Ref.\ \citeo{ourpre} that this 
equality between the two
surface exponents is valid everywhere close to the transition line,
and not just where $D=\lambda+\alpha$. 

At the line $D=\lambda+\alpha$,
it is now straightforward to derive a second relation
\begin{equation}
\beta^{\rm Sp}_{1,{\rm seed}}=\beta^{\rm O}_{1,{\rm
dens}} . \label{duality2}
\end{equation}
One simply starts off
with the quantum Hamiltonian $\tilde{\cal H}$ and then
follows the same steps as above. $\tilde{\cal H}$ can then be mapped
back onto the starting Hamiltonian ${\cal H}$, meaning that the
transformation is actually a duality transformation. A relation
like that in Eq.~(\ref{equiv}) can then be derived, giving the above result.

In summary, in $d=1$ and at the particular line in parameter space
$D=\lambda+\alpha$, we have mapped BARW at the
special transition onto BARW at the ordinary transition
(and vice-versa), a rather non-trivial procedure
reminiscent of a related transformation in the 
Ising model. Unfortunately, the results (\ref{duality1})
and (\ref{duality2}) are only derived for one line in parameter
space and we have to rely on universality in order to claim that they are
valid elsewhere close to the transition line.

\break

\section{Numerical Methods}
\label{sec:numres}
\noindent
We now discuss the various numerical methods which can be used
to obtain precise estimates for the critical exponents for boundary
DP and BARW.
The values of the exponents are listed in Section~\ref{simreslist}.

\subsection{Extraction of exponents from Monte-Carlo simulations}
\noindent
In this subsection we review how the exponents are extracted from
Monte-Carlo simulation data for the DP and DP2 models with walls
in $d=1$ and $d=2$.\cite{lauritsen-etal,dp-wall-edge,ourprl,ourpre}
Numerics have also been carried out for BARW with a wall in $d=1$,
with results in good agreement with boundary DP2
simulations.\cite{ourpre} The most efficient way of extracting
information on boundary DP and DP2 is to perform simulations
at criticality, starting from an inital seed next to the wall.
Measurements can then be taken of the survival
probability $P_1(t)$, the activity
in the bulk $N_1(t)$ and at the wall $N_{1,1}(t)$, the average
spread of the cluster $\left< x^2(t) \right>$,
and the probability $p_1(s)$ to have a cluster of
size (mass) $s$.\cite{grassberger-torre} 
Furthermore, by averaging over surviving clusters only (denoted by an
over-line), measurements can be taken of the surviving bulk activity
$\overline{N_{1}}(t)$ 
and the surviving wall activity $\overline{N_{1,1}}(t)$.
Although carrying out these simulations is straightforward, extracting
the exponents discussed in the previous sections requires some further
analysis. We will now review the scaling theory which allows these
connections to be made.\cite{exponen} Note that all the
relations given below are valid for the DP ordinary transition and for both 
the IBC (special) and RBC (ordinary) DP2 transitions, and hence
these labels will be suppressed from now on.

The probability for a cluster grown from a seed on the wall
still to be alive at time $t$ is given by Eqs.~(\ref{eq:P_1(t,Delta)})
and (\ref{barwsurvwall}). At criticality ($\Delta = 0$) it has the
following behavior 
\begin{equation}
        P_1(t) = t^{-\delta_{1, \rm seed}} f(t/t_c) , \qquad
\delta_{1, \rm seed} = \beta_{1, \rm seed}/\nu_{\parallel} ,
                                         \label{eq:P_1(t)}
\end{equation}
where the scaling function $f$ depends on the cutoff
$t_c \sim \xi_\parallel \sim |\Delta|^{-\nu_\parallel}$.
Hence, the probability of growing a cluster which lives exactly $t$
time steps behaves as $p_1(t)\sim t^{-1 - \delta_{1, \rm seed}}$.

The average number of active sites at criticality,
averaged over all clusters, is obtained by integrating 
the density (\ref{DPansatz_rho_wall}) and (\ref{ansatz_rho_wall}) over
space, giving
\begin{equation}
        N_1(t) \sim t^{\kappa_1}, \qquad
        \kappa_1 = d\chi - \delta_{\rm dens} - \delta_{1, \rm seed}, 
                                        \label{eq:N_1(t)}
\end{equation}
where we have introduced the cluster envelope or ``roughness'' exponent
$\chi=\nu_\perp/\nu_\parallel$ ($\equiv 1/z$), and the notation
$\delta_{\rm dens} = \beta_{\rm dens}/\nu_{\parallel}$.
Note that the scaling relation in Eq.~(\ref{eq:N_1(t)}) corresponds to
the hyperscaling relations (\ref{DPsurf_hyperscaling}) and
(\ref{surf_hyperscaling}), a fact which follows from 
$\langle s_1(t) \rangle = \int_0^t dt' \, N_{1}(t')$, and the relation
$\gamma_1 = \nu_\parallel (1+\kappa_1)$.
By integrating the density on the wall (\ref{DPansatz_rho_wall_wall})
and (\ref{ansatz_rho_wall_wall}), we obtain the average number of
active sites at criticality on the wall
\begin{equation}
        N_{1,1}(t) \sim t^{\kappa_{1,1}}, \qquad
        \kappa_{1,1} 
             = (d-1)\chi - \delta_{1, \rm dens} - \delta_{1, \rm seed}, 
                                        \label{eq:N_1,1(t)}
\end{equation}
where
$\delta_{1, \rm dens} = \beta_{1, \rm dens}/\nu_{\parallel}$.
Note also that the scaling relation in Eq.~(\ref{eq:N_1,1(t)})
corresponds to the hyperscaling relations
(\ref{DPsurf_hyperscaling_wall_wall}) and
(\ref{surf_hyperscaling_wall_wall}) at criticality,
since $\langle s_{1,1}(t) \rangle = \int_0^t dt' \, N_{1,1}(t')$, and 
$\gamma_{1,1} = \nu_\parallel (1+\kappa_{1,1})$.

Alternatively, by averaging only over clusters
which survive to infinity (denoted by an over-line), we obtain
\begin{equation}
        \overline{N_{1}}(t) \sim t^{\overline{\kappa_1}},\qquad
        \overline{\kappa_1} = d\chi - \delta_{\rm dens}.
                                        \label{eq:overline-N_1}
\end{equation}
The activity on the wall for surviving clusters reads
\begin{equation}
        \overline{N_{1,1}}(t) \sim t^{\overline{\kappa_{1,1}}} ,\qquad 
        \overline{\kappa_{1,1}} = (d-1)\chi - \delta_{1, \rm dens}.
                                        \label{eq:overline-N_1,1}
\end{equation}
Simulations in $d=1$ thus directly yield
$\delta_{1, \rm dens}$.  The average position of activity
follows from Eqs.~(\ref{DPansatz_rho_wall}) and
(\ref{ansatz_rho_wall}), giving $\left< x^2 \right> 
\sim  t^{2\chi}$, where $x$ is the distance from the seed and the
average is taken over all active points at a given time.

Additional numerical data was obtained in
Refs.\ \citeo{ourprl,ourpre} by considering the
cluster size distributions at criticality. In the bulk the typical
cluster size $s_c$ of finite clusters scales as volume times density,
i.e.,
\begin{equation}
        s_c \sim \xi_\perp^d \xi_\parallel n(\Delta) \sim
        |\Delta|^{-1/\sigma}, \qquad
        1/\sigma = d\nu_\perp + \nu_\parallel - \beta_{\rm dens}.
                                \label{eq:s-typical}
\end{equation}
{}From the lifetime survival distribution (\ref{eq:P_1(t)}), it is
then straightforward to obtain the probability $P_1(s)$ to have a
cluster of size larger than $s$, for clusters started from a seed on
the wall. Using the fact that the lifetime
is set by the parallel correlation length,
$t\sim \xi_\parallel \sim |\Delta|^{-\nu_\parallel}$, we see that
the typical cluster size and lifetime are connected by
$s \sim t^{1/\nu_\parallel\sigma}$.
Hence we obtain $P_1(s) \sim P_1(t\sim s^{\nu_\parallel\sigma})
\sim s^{-\beta_{1,{\rm seed}} \sigma}$.
Thus, we eventually obtain the probability $p_1(s)$ to have
a cluster of exactly size $s$, $p_1(s)= - dP_1(s)/ds$, with the
result
\begin{equation}
   p_1(s) = s^{-\mu_1} g(s/s_c)  ,\qquad
   \mu_1 = 1 + \frac{\beta_{1, \rm seed}}
               {d \nu_\perp + \nu_\parallel - \beta_{\rm dens}}.
                                \label{eq:p_1(s)}
\end{equation}
Similarly, the cluster size distribution on the wall due to a 
seed located at the wall can also be obtained.
In this case the typical cluster size of finite clusters is
\begin{equation}
        s_{\rm wall, c} \sim \xi_\perp^{d-1} \xi_\parallel n_1(\Delta)
        \sim |\Delta|^{-1/\sigma_1} , 
                                \label{eq:s_wall-typical}
\end{equation}
where the cutoff exponent is
\begin{equation}
        1/\sigma_1 = (d-1)\nu_\perp + \nu_\parallel - \beta_{1, \rm
        dens} . 
                                \label{eq:1/sigma_1}
\end{equation}
The resulting distribution reads
\begin{equation}
        p_{1,1}(s_{\rm wall}) 
		= s_{\rm wall}^{-\mu_{1,1}} f(s_{\rm wall}/s_{\rm wall, c} )  ,
                                \label{eq:p_1,1(s)}
\end{equation}
with
\begin{equation}
        \mu_{1,1} = 1 + \frac{\beta_{1, \rm seed}}
             {(d-1) \nu_\perp + \nu_\parallel - \beta_{1, \rm dens}}.
                                \label{eq:mu_1,1}
\end{equation}



With the above scaling relations it is now straightforward to extract
all the exponents from the numerical data. Values for the exponents
obtained in this way are listed in Section~\ref{simreslist}. 
An additional interesting exponent relation for the RBC
transition for DP2 can be obtained by assuming that the survival
probability is dominated by the return to the wall of the
cluster-envelope. This leads to\cite{lauritsen-etal}  
\begin{equation}
	\delta^{\rm RBC}_{1, \rm seed} = 1-\chi ,
				\label{eq:1-chi}
\end{equation}
in agreement with the simulation results for the RBC (see
Section~\ref{simreslist}). 
Qualitatively, this means that the $I_2$ regions located at the wall
determine the scaling since they can only disappear when the activity
returns to the wall. Note that a relation of this kind clearly fails
for the IBC transition. Furthermore, if the cluster lifetime is
defined to be the return time of the cluster-envelope (i.e.\ the return
time of the rightmost active site) to the initial point, then
we expect clusters defined in this way to have a lifetime distribution 
exponent $\delta_{1,{\rm seed}}$ given by Eq.~(\ref{eq:1-chi}). This
prediction is in agreement
with the simulations in Ref.\ \citeo{hwang-etal:1998},
where various models in the DP and BARW classes were studied
with cluster lifetimes defined in the way described above.
For further information on the DP2 boundary exponents, including
inequalities for the $\beta_{1,{\rm seed}}$ and $\beta_{1,{\rm dens}}$
exponents, we refer to Ref.\ \citeo{ourpre}.

\subsection{Series expansions}
\noindent
In the previous subsection we reviewed the scaling relations used to 
extract exponents from Monte-Carlo simulations. These simulations provide 
fairly accurate exponent estimates for both the DP and BARW classes 
(see the next subsection). However, for the case of DP in $d=1$
(but curiously {\em not} for BARW in $d=1$\cite{jensen:1997}) this
level of accuracy can be far surpassed by using  
the method of series expansions.\cite{essam-etal:1996,newjensen} 
With uncertainties in the seventh or eighth digit, these exponent 
values are very reliable and useful for testing conjectures and scaling
laws. For example, most of the DP exponents in $d=1$ are now known not
to lie close to any simple rational numbers. In fact, it seems more
likely that the exponents are irrational. However, one important
exception is the exponent $\tau_1$ governing the mean cluster
lifetime in the presence of a wall (see 
Eq.\ (\ref{taudef1})). This exponent has been conjectured to
equal unity,\cite{essam-etal:1996} although this has now
been challenged by the estimate $\tau_1 = 1.00014(2)$
(see below).\cite{newjensen} 

The idea behind series expansions in DP is to find an effective algorithm
for calculating the expansion coefficients of moments of the
pair-connectedness (the probability that the site $x$ at time
$t$ is occupied given a seed at the origin at $t=0$). By analyzing
these expansions, critical parameters of the model can be obtained,
including estimates for $p_c$, and the exponents $\gamma$, $\nu_\perp$,
and $\nu_\parallel$. 
These series expansions can be generated by transfer matrices that
relate states at time $t$ to those at one time step later.
At any time $t$ this gives polynomials in the percolation probability $p$.
Since the difficulty in generating these series expansions is of
exponential complexity, a large amount of effort has been devoted
to keeping the computational effort as small as possible (see Ref.\
\citeo{jensen:1999} and references therein). Once a series expansion
is derived, it is then analyzed by using differential approximants.
The more terms that can be computed in the series, the more accurately
the exponents can be estimated.


\subsection{Exponent Values}
\label{simreslist}
\noindent
In this section we list the best exponent values currently available
for the surface and bulk exponents for DP and BARW. As we have
mentioned, for DP in $d=1$ the best results are from series
expansions; in all other cases we give Monte-Carlo data.
Note that various attempts have been made to fit these exponent values
with rational numbers; further details can be found in
Refs.\ \citeo{essam-etal:1996,ourpre}.

\begin{table}[bhtbp]
\tcaption{Critical exponents for DP in $d=1$ and $d=2$ in
        the bulk and for the ordinary transition at the boundary.
	}
\vspace*{5mm}
\centerline{\footnotesize\smalllineskip
\begin{tabular}{l|l|l|c}
              & \makebox[20mm]{$d=1$} & \makebox[16mm]{$d=2$} 
              & \makebox[16mm]{Mean Field} \\
\hline
$\delta_{\rm dens}=\delta_{\rm seed}$    
                       & ~0.159 464(6) & ~0.450(1)   &   1 \\
$\delta_{1, \rm dens}=\delta_{1,\rm seed}$  
                       & ~0.423 17(2)   & ~0.82(4)   &  3/2  \\
$\kappa$               & ~0.313 686(8)  & ~0.230(1)  &   0  \\ 
$\kappa_1$             & ~0.049 98(2)   & ~-0.13(1)  & -1/2 \\ 
$\overline{\kappa}=\overline{\kappa_1}$  
                       & ~0.473 150(10)  & ~0.682(1)  & 1 \\
$\mu$                  & ~1.108 247(10) & ~1.268(8)  &  3/2  \\
$\mu_1$                & ~1.287 25(2)   & ~1.49(8)   &  7/4  \\
$2\chi$                & ~1.265 226(13) & ~1.132(7)  &  1  \\
\hline
$\beta_{\rm dens}=\beta_{\rm seed}$     
                       & ~0.276 486(8) & ~0.583(4)   &   1 \\
$\beta_{1, \rm dens}=\beta_{1,\rm seed}$   
                       & ~0.733 71(2)   & ~1.07(5)   &  3/2  \\
$\tau$                 & ~1.457 362(14)  & ~0.711(7)  &    0  \\
$\tau_1$               & ~1.000 14(2)   & ~0.26(2)   &    0  \\
$\gamma$               & ~2.277 730(5)  & ~1.593(7)  &    1 \\
$\gamma_1$             & ~1.820 51(1)   & ~1.05(2)   &  1/2 \\
\end{tabular}}
\label{table-dp}
\end{table}

\begin{table}[htbp]
\tcaption{Critical exponents for the BARW class measured from
        simulations of DP2 in $d=1$.   
  	}
\vspace*{5mm}
\centerline{\footnotesize\smalllineskip
\begin{tabular}{l|l|l|l}
              & \makebox[16mm]{$d=1$} & \makebox[16mm]{$d=1$ (IBC)} 
              & \makebox[16mm]{$d=1$ (RBC)} \\
\hline
$\delta_{\rm dens}$
              & ~0.287(5) &  ~0.288(2) & ~0.291(4)\\
$\delta_{\rm seed}$
              & ~0.290(5) &  & \\
$\delta_{1,\rm seed}$
              &           &  ~0.641(2) & ~0.426(3) \\
$\delta_{1, \rm dens}$
              &           &  ~0.415(3) & ~0.635(2) \\
$\kappa$      & ~0.000(2) &              &        \\
$\kappa_1$    &           & ~-0.354(2)~  & ~-0.141(2)~\\
$\overline{\kappa} = \overline{\kappa_1} $
              & ~0.288(5)  &  ~0.287(2)  & ~0.285(2)\\
$\mu$         &  ~1.225(5) &             &    \\
$\mu_1$       &            &  ~1.500(3)  & ~1.336(3) \\
$\mu_{1,1}$   &  ~1.408(5) &  ~2.05(5)   & ~2.15(5)  \\
$2\chi$       &  ~1.150(5) &   ~1.150(3) & ~1.152(3) \\
\hline
$\beta_{\rm dens}$
              & ~0.922(5) &  ~0.93(1)  & ~0.94(2) \\
$\beta_{\rm seed}$
              & ~0.93(5)  &  & \\
$\beta_{1,\rm seed}$
              &           &  ~2.06(2)  & ~1.37(2)  \\
$\beta_{1, \rm dens}$
              &           &  ~1.34(2)  & ~2.04(2)  \\
$\tau$        &  ~2.30(3) &            &   \\
$\tau_1$      &           &  ~1.16(4)  & ~1.85(4)   \\
$\gamma$      &  ~3.22(5) &            &   \\
$\gamma_1$    &           &  ~2.08(4)  & ~2.77(4)  \\
$\gamma_{1,1}$&  ~1.38(3) &  ~~($<0$)  & ~~($<0$)  \\
\end{tabular}}
\label{table-exp1}
\end{table}

\break

\section{Other Directions}
\label{otherwork}
\vglue 0pt
\noindent
In this section we briefly mention some other nonequilibrium systems where
boundary effects have been analyzed.
Additional studies, which we will not discuss, include (i) the
connection between surface DP and local persistence;\cite{hayehari}
(ii) $d=1$ density matrix renormalization group calculations
of some reaction-diffusion processes;\cite{nancy}
(iii) active, but slanted, walls in DP, which give rise 
to a ``curtain'' of activity whose width is given by an
angle-dependent correction to bulk DP;\cite{seattle} and,
finally, (iv) Monte-Carlo simulation studies of rigidity percolation
with and without walls, which have shown the model to belong to the
DP universality class.\cite{menezes-moukarzel}

\subsection{Compact DP}
\noindent
By forcing a DP cluster to be dense one obtains a different universality 
class called Compact Directed Percolation.\cite{domany1} This actually 
corresponds to the special case $q=1$ of the Domany-Kinzel model in 
Figures~\ref{DPlattice}~and~\ref{DPrules}. The constraint $q=1$ forces 
all interior sites and/or bonds to be active, whereas the size of the cluster
is controlled by $p$, which now only affects the boundaries of the cluster.
As the dynamics of the cluster is entirely controlled by
the dynamics of its surface, the problem is greatly 
simplified. For instance, in $d=1$, the extinction of a
compact DP cluster can be viewed as the probability for a pair of random
walkers to coincide. It is therefore no surprise that compact DP can be
exactly solved in $d=1$ and that all the exponents are simple
integers.\cite{domany1}
For bulk compact DP in $d=1$, one has $\beta_{\rm seed} = 1$,
$\nu_\parallel = 2$, $\nu_\perp = 1$, and $\gamma = 2$.

By introducing a wall in compact DP, the survival probability is
altered and one obtains surface critical exponents
just as for DP. With an {\em inactive wall}, the cluster is free to approach 
and leave the wall, but not cross. For $d=1$, this gives rise
to $\beta_{1,\rm seed} = 2$, i.e., twice as big as $\beta_{\rm seed}$ for the 
bulk.\cite{bidaux,lin} On the 
other hand, for an {\em active wall}, the cluster is stuck to the wall and
therefore described by a single random walker for $d=1$.
By reflection in the wall,
this may be viewed as {\em symmetric} compact DP which has the same 
$\beta_{\rm seed}$ as normal compact DP, 
giving $\beta_{1,\rm seed} = 1$.\cite{essamtanlakishani} 

\subsection{Backbone and red bonds in DP}
\noindent
In Ref.\ \citeo{lauritsen-etal} the so-called backbone and red bonds
of $d=1$ DP clusters have been investigated.
The backbone is obtained from the infinite cluster by
removing all dangling ends. Thus the backbone consists of precisely 
those bonds which would be occupied by both the time-directed DP
process and its reverse time-directed process. It then follows that
the backbone density $|\Delta|^{\beta^{BB}}$ is described by
the exponent $\beta^{BB} = 2\beta$. Numerically, the backbone
dimension on the wall was measured with the result $D_{1}^{BB} = 0.16
\pm 0.01$. Using the scaling relation $D_{1}^{BB} = 1 -
\beta_{1}^{BB}/\nu_\parallel$ it then follows that
$\beta_{1}^{BB} = 1.46$ in good agreement with $\beta_{1}^{BB}=2\beta_1$,
cf.\ the result for the bulk.

On the backbone one can identify so-called red bonds\cite{stanley}
that, if one is cut, divide the cluster into two parts.
A renormalization group argument\cite{cognilio}
(see also Ref.\ \citeo{huber}) yields that the number of red bonds up to
time $t$ scales as
\begin{equation}
        N_R(t) \sim t^{1/\nu_\parallel}.
                \label{eq:red-bonds}
\end{equation}
Reference \citeo{lauritsen-etal} measured
the scaling of red bonds for DP with a wall
and obtained results in complete agreement with Eq.~(\ref{eq:red-bonds}).
In addition the scaling
of red bonds along a longitudinal cut for DP with no wall was measured
with the result $N_{R}^{cut}(t) \sim t^{-0.04 \pm 0.02}$.
This is in accordance with the expected behavior
$N_{R}^{cut}(t) \sim t^{1/\nu_\parallel - \nu_\perp/\nu_\parallel}
\sim t^{-0.056}$,
where the extra factor originates from the scaling of the width
$\xi_\perp\sim t^{\nu_\perp/\nu_\parallel}$ of the cluster.
Finally, the scaling of boundary red bonds for
DP with the wall was measured, where the behavior
$N_{R,1}(t) \sim t^{-0.60 \pm 0.1}$ was found.

\subsection{Invasion Percolation}
\noindent
Finally, we briefly discuss a slightly different system,
namely invasion percolation (IP), which
is a model for the growth of infinite percolation clusters
at criticality.\cite{invperc} In $d=2$ 
the fractal dimension of IP clusters is $D=2-\beta/\nu =
2 - (5/36) / (4/3) = 91/48 \approx 1.90$, where the known exact 
values for percolation in $d=2$ have
been used (see, e.g., Ref.\ \citeo{surfperc}).  

In Ref.\ \citeo{cafiero},
boundary effects in the growth of $d=2$ invasion percolation clusters
were studied.  Numerically, near a wall it was found that the 
fractal dimension of IP clusters was $D_s=1.67 \pm 0.03$.
Using the scaling theory for boundary nonequilibrium systems,
it follows that the activity on the wall
has the fractal dimension $D_1 = 1 - \beta_1 / \nu$,
where $\beta_1$ is the surface exponent for $d=2$ percolation.
Using $\beta_1=4/9$, we obtain that $D_1=2/3$.
The fractal dimension of parts of the cluster close to the wall
is thus predicted to be $D_1 + 1 = 5/3$ in excellent agreement
with the simulation result.

\section{Summary\label{summary}}
\vglue 0pt
\noindent
In this review we have outlined the boundary critical behavior
of some nonequilibrium systems, with a particular focus on the
directed percolation and 
branching-annihilating random walk universality classes.
Through the use of a variety of theoretical and numerical techniques,
including mean field, scaling and field theories, exact solutions,
Monte-Carlo simulations, and series expansions, a considerable
amount of progress has been made in understanding the boundary
critical properties of these models. Nevertheless some important
problems remain open and we wish to conclude this review with a brief
list of some of these remaining questions.  

\

\noindent
1. Due to the presence of a second critical dimension $d_c'$,
the BARW field theory in $d=1$ is not well-controlled either
at the boundary or in the bulk. An improved theory would be highly
desirable in confirming the results provided by exact solutions
and Monte-Carlo simulations.

\

\noindent
2. Although we have not discussed it in this review, the boundary
critical properties of {\em dynamical percolation} form another
interesting case (for $d\geq 2$).\cite{grassberger-3d} In this
universality class the development of an appropriate boundary
field theory, and the calculation of boundary exponents using
epsilon expansion techniques, has not yet been attempted. 

\

\noindent
3. The intriguing question of whether $\tau_1=1$ for the DP
ordinary transition in $d=1$ remains unsolved. No theoretical
explanation has emerged for why $\tau_1$ should equal unity, and, in fact,
the latest series expansions yield a value for $\tau_1$ very slightly
away from $\tau_1=1$. Hence the possibility of numerical coincidence,
where $\tau_1$ just happens to lie very close to unity, cannot be ruled out. 

\

\noindent
4. Most of the numerical work so far has focused on the ordinary
transition. Particularly for the case of DP, it would be useful to
measure the exponents at the extraordinary and special transitions for $d>1$.

\nonumsection{Acknowledgements}
\noindent
Part of this work was carried out while M.H. and K.B.L. were at the
Niels Bohr 
Institute. P.F. acknowledges support from the Swedish Natural Science
Research Council. M.H. acknowledges support from the CATS group at the
Niels Bohr Institute, from the NSF through the
Division of Materials Research and from NSERC of
Canada. K.B.L. acknowledges support from the Carlsberg Foundation. 

\nonumsection{References}


\begin{thebibliography}{000}
\bibitem{diffprl}
	Note that the convention used for $\Delta$ here is different
	from that in Ref. \ \citeo{ourprl}.
\bibitem{mistake}
        Note that the mean field surface density exponent
        $\beta_{1,{\rm dens}}^{\rm IBC}$ quoted in
        footnote [22] of Ref.\ \citeo{ourprl} is incorrect.
\bibitem{exponen}
        Note that some of our exponent definitions in this section differ  
        from those sometimes found in the literature. In particular
        our $\kappa$ exponents are sometimes called $\eta$, and the
        dynamic exponent $z$ is sometimes redefined as $z\to 2/z$.
\end{thebibliography}
\end{document}